\documentclass[aps,prd,twocolumn,amsfonts,amssymb,amsmath,showpacs,groupedaddress,letterpaper,floatfix]{revtex4}

\usepackage{graphicx}
\usepackage{dcolumn}
\usepackage{color}

\newcommand{\BaBarYear}{09}
\newcommand{\BaBarNumber}{004}
\newcommand{\SLACPubNumber}{13735}
\newcommand{\BaBarType}{PUB}

\def\optbar#1{\vbox{\ialign{##\crcr\hfil${\scriptscriptstyle(}\mkern -1mu
    \vrule height 1.2pt width 3pt depth -.8pt {\scriptscriptstyle)}$
    \hfil\crcr \noalign{\kern-1pt\nointerlineskip}$\hfil\displaystyle{#1}\hfil$\crcr}}}
\input{pubboard/babarsym}
\newcommand{\resultfitchisqmx}{10.9}
\newcommand{\resultfitchisqnx}{8.2}

\newcommand{\gevsq}{\ensuremath{\mathrm{\,Ge\kern -0.1em V^2}}\xspace}
\newcommand{\gevcube}{\ensuremath{\mathrm{\,Ge\kern -0.1em V^3}}\xspace}
\newcommand{\gevtothe}[1]{\ensuremath{\mathrm{\,Ge\kern -0.1em V^{#1}}}\xspace}

\def\X       {\ensuremath{X}\xspace}
\def\Xc      {\ensuremath{X_{c}}\xspace}

\def\Breco   {\ensuremath{\B_\mathrm{tag}}\xspace}

\def\Brecoil   {\ensuremath{\B_\mathrm{\rm recoil}}\xspace}
\def\ellm    {\ensuremath{\ell^{-}}\xspace}
\def\ellp    {\ensuremath{\ell^{+}}\xspace}
\def\porantip    {\kern 0.18em\optbar{\kern -0.40em p}{}\xspace}
\def\D       {\ensuremath{D}\xspace}
\def\DorDstar    {\ensuremath{\D^{\left(*\right)}}\xspace}
\def\DbarorDbarstar    {\ensuremath{\Dbar^{\left(*\right)}}\xspace}

\newcommand{\semilepXu} { \ensuremath{\Bbar \rightarrow \X_{u} \ellm \nub }}
\newcommand{\semilepXc} { \ensuremath{\Bbar \rightarrow \X_{c} \ellm \nub }}

\newcommand{\semilepXce} { \ensuremath{\Bbar \rightarrow \X_{c} \en \nub }}

\newcommand{\semilepBpXc} { \ensuremath{\Bp \rightarrow \X_{c} \ellp \nu }}
\newcommand{\semilepBzXc} { \ensuremath{\Bz \rightarrow \X_{c} \ellp \nu }}

\newcommand{\semilepD} { \ensuremath{\Bbar \rightarrow \D \ellm \nub }}
\newcommand{\semilepDstar} { \ensuremath{\Bbar \rightarrow \Dstar \ellm \nub }}

\newcommand{\semilepDstarstar} { \ensuremath{\Bbar \rightarrow \Dstarstar \ell \nub }}
\newcommand{\semilepNreso} { \ensuremath{\Bbar \rightarrow \DorDstar \pi \ell \nub }}

\newcommand{\semilepDoneprime}{\ensuremath{\Bbar \rightarrow \D_1' \ellm \nub}\xspace}
\newcommand{\semilepDone}{\ensuremath{\Bbar \rightarrow \D_1 \ellm \nub}\xspace}
\newcommand{\semilepDtwostar}{\ensuremath{\Bbar \rightarrow \D_2^* \ellm \nub}\xspace}
\newcommand{\semilepDzstar}{\ensuremath{\Bbar \rightarrow \D_0^* \ellm \nub}\xspace}

\newcommand{\BtoXsGamma} { \ensuremath{\Bbar\rightarrow \X_{s} \g }}
\newcommand{\Gammasl}    {\ensuremath{\Gamma_{\mathit{SL}}}\xspace}

\newcommand{\Dstarstar}{\ensuremath{D^{**}}\xspace}

\def\photos      {\mbox{\tt PHOTOS}\xspace}
\def\geantf      {\mbox{\sc GEANT4}\xspace}

\newcommand{\mxmom}[1]  { \ensuremath{\langle m_{\X}^{#1} \rangle} }

\newcommand{\mxmomreco}[1]  {\ensuremath{\langle m_{\X,reco}^{#1}\rangle }}
\newcommand{\mxmomcalib}[1]  {\ensuremath{\langle m_{\X,calib}^{#1}\rangle }}

\newcommand{\mxmomtruecut}[1]  {\ensuremath{\langle m_{\X,true}^{#1}\rangle }}
\newcommand{\mx}  {\ensuremath{m_{\X}\xspace}}
\newcommand{\mxtrue}  {\ensuremath{m_{\X,true}}}

\newcommand{\mxreco}  {\ensuremath{m_{\X,reco}}}
\newcommand{\mxrecoi} {\ensuremath{m_{\X,reco,i}}}

\newcommand{\mxcalibi}  {\ensuremath{m_{\X,calib,i}}}

\newcommand{\elmom}[1]  { \ensuremath{\langle E_{\ell}^{#1}\rangle } }
\newcommand{\egammamom}[1]  { \ensuremath{\langle E_{\gamma}^{#1}\rangle } }

\newcommand{\Ex}  {\ensuremath{E_{\X}\xspace}}
\newcommand{\nx}  {\ensuremath{n_{\X}^{2}\xspace}}
\newcommand{\nxfour}  {\ensuremath{n_{\X}^{4}}}
\newcommand{\nxsix}  {\ensuremath{n_{\X}^{6}}}
\newcommand{\moment}[1]{\ensuremath{\langle {#1}\rangle }}

\newcommand{\nxn}{\ensuremath{n_{\X}^{k}}}

\newcommand{\Pmissfourmom}     { \ensuremath{ P_\mathrm{miss} } }

\newcommand{\epmiss}     { \ensuremath{E_\mathrm{miss} - c p_\mathrm{miss}} }
\newcommand{\epmissabs}  { \ensuremath{\left| \epmiss \right|} }
\newcommand{\emiss}      { \ensuremath{E_\mathrm{miss}} }
\newcommand{\pmiss}      { \ensuremath{p_\mathrm{miss}} }

\newcommand{\plep}         { \ensuremath{p_{\ell}^{*}} }
\newcommand{\plepe}         { \ensuremath{p_{e}^{*}} }
\newcommand{\plmin}      { \ensuremath{p_{\ell,\mathrm{min}}^{*}} }
\newcommand{\plgeq}[1]   { \ensuremath{\plep \geq {#1} \gevc }}
\newcommand{\plbin}[2]   { \ensuremath{ {#1} \leq \plep <  {#2} \gevc }}
\newcommand{\plcut}      { \ensuremath{p_{\ell,\mathrm{min}}^{*}} }

\newcommand{\Egammacut}      { \ensuremath{E_{\gamma,\mathrm{min}}} }
\newcommand{\Egammageq}[1]      { \ensuremath{E_{\gamma} \geq {#1} \gev} }

\newcommand{\brf}        { \ensuremath{\mathcal{B}}}

\newcommand{\MultX}      { \ensuremath{N_{\Xc}} }

\def\mesmax     {\mbox{$m_{\mathrm{ES,max}}$}\xspace}

\newcommand{\Ctrue} { \ensuremath{\mathcal{C}_{\mathit{true}}}}
\newcommand{\Ccalib} { \ensuremath{\mathcal{C}_{\mathit{cal}}}}

\def\wi         { \ensuremath{w_{i}}\xspace }

\def\cov        { \ensuremath{\mathcal{C}}}

\newcommand{\mb}         { \ensuremath{m_{\b}}\xspace }
\newcommand{\mc}         { \ensuremath{m_{\c}}\xspace }
\newcommand{\mbbar}         { \ensuremath{\overline{m}_{\b}}\xspace }
\newcommand{\mcbar}         { \ensuremath{\overline{m}_{\c}}\xspace }

\newcommand{\mupi}      { \ensuremath{\mu_{\pi}^{2}}\xspace }
\newcommand{\muG}      { \ensuremath{\mu_{G}^{2}}\xspace }
\newcommand{\rhoD}      { \ensuremath{\rho_{D}^{3}}\xspace }
\newcommand{\rhoLS}      { \ensuremath{\rho_{\mathit{LS}}^{3}}\xspace }

\newcommand{\covhqe}{ \ensuremath{\cov_{\mathrm{HQE}}}} 
\newcommand{\covexp}{ \ensuremath{\cov_{\mathrm{exp}}}} 
\newcommand{\covtot}{ \ensuremath{\cov_{\mathrm{tot}}}}

\newcommand{\etag}{\ensuremath{e_\mathrm{tag}}\xspace}
\newcommand{\esig}{\ensuremath{e_\mathrm{sig}}\xspace}

\newcommand{\epmisslow} {-0.2}
\newcommand{\epmissup}  {0.3}
\newcommand{\emisslow}  {0}
\newcommand{\pmisslow}  {0}
\newcommand{\percentResidualBG}{10\,\%\xspace}
\newcommand{\percentCombBG}    {12\,\%\xspace}
\newcommand{\percentBG}        {22\,\%\xspace}
\newcommand{\NsigFirst}  {$10,053 \pm 142$}
\newcommand{\NsigSecond} {$1,626 \pm 52$}
\newcommand{\resolNx}       {1.05\,\gevsq}
\newcommand{\biasNx}        {-0.11\,\gevsq}
\newcommand{\resolNxbefore} {1.17\,\gevsq}
\newcommand{\biasNxbefore}  {-0.37\,\gevsq}

\long\def\inst#1{\par\nobreak\kern 4pt\nobreak
    {\it #1}\par\ubboard/vskip 10pt plus 3pt minus 3pt}

\begin{document}

\begin{flushleft}
\babar-\BaBarType-\BaBarYear/\BaBarNumber \\
SLAC-PUB-\SLACPubNumber \\
\end{flushleft}

\title{\large \bf
\boldmath
Measurement and Interpretation of Moments in Inclusive
Semileptonic Decays $\semilepXc$
} 

%
\author{B.~Aubert}
\author{Y.~Karyotakis}
\author{J.~P.~Lees}
\author{V.~Poireau}
\author{E.~Prencipe}
\author{X.~Prudent}
\author{V.~Tisserand}
\affiliation{Laboratoire d'Annecy-le-Vieux de Physique des Particules (LAPP), Universit\'e de Savoie, CNRS/IN2P3,  F-74941 Annecy-Le-Vieux, France}
\author{J.~Garra~Tico}
\author{E.~Grauges}
\affiliation{Universitat de Barcelona, Facultat de Fisica, Departament ECM, E-08028 Barcelona, Spain }
\author{M.~Martinelli$^{ab}$ }
\author{A.~Palano$^{ab}$ }
\author{M.~Pappagallo$^{ab}$ }
\affiliation{INFN Sezione di Bari$^{a}$; Dipartimento di Fisica, Universit\`a di Bari$^{b}$, I-70126 Bari, Italy }
% ...
\collaboration{The \babar\ Collaboration}
\noaffiliation

\date{\today}

\begin{abstract}
We present results for the moments of observed spectra in inclusive semileptonic
$\B$-meson decays to charm hadrons $\Bbar \to X_c \ell^{-} \bar{\nu}$. 
Moments of the hadronic-mass and the combined mass-and-energy spectra for different 
minimum electron or muon momenta between $0.8$ and $1.9 \gevc$ are obtained 
from a sample of $232 \times 10^{6}$ $\Upsilon(4S)\to \B \Bbar$ events, 
collected with the $\babar$ detector at the PEP-II asymmetric-energy $\B$-meson factory at SLAC.
We also present a re-evaluation of the moments of electron-energy spectra and partial decay fractions 
$\BR(\Bbar \to X_c \en \bar{\nu})$ for minimum electron momenta between $0.6$ and $1.5 \gevc$
based on a sample of $51 \times 10^{6}$ $\FourS \to \BB$ events. 
The measurements are used for the extraction of the total decay fraction, 
the Cabibbo-Kobayashi-Maskawa (CKM) matrix element $\Vcb$, the quark masses
$\mb$ and $\mc$, and four heavy-quark QCD parameters in the framework of a 
Heavy Quark Expansion (HQE). We find
$\BR(\Bbar \to X_c \ell^{-} \bar{\nu}) = (10.64 \pm 0.17 \pm 0.06)\%$ and
$\Vcb = (42.05 \pm 0.45 \pm 0.70) \times 10^{-3}$.
\end{abstract}

\pacs{12.15.Ff, 12.15.Hh, 13.25.Hw, 13.30.Ce}

\maketitle

\setcounter{footnote}{0}

\section{Introduction}
\label{sec:introduction}

The Standard Model of particle physics (SM) contains a large number of free
parameters which can only be determined by experiment. Precision measurements
of all of these parameters are essential for probing the validity
range of the model by comparing many other precision measurements to
SM calculations. Three of the parameters, the Cabibbo-Kobayashi-Maskawa (CKM)
matrix element \Vcb \cite{1963:CabibboCKM,1973:KobayashiCKM} and the 
heavy quark masses $m_b$ and $m_c$, can be related via Operator Product Expansions (OPE) 
to moments and rates of inclusive distributions in semileptonic $\B$ meson 
decays, $\semilepXc$ \cite{ChargeConjugation}, 
and rare $\B$-meson decays, $\BtoXsGamma$, where $X_{c}$ and
$X_{s}$ denote the hadronic systems with
charm and strangeness in these final states, respectively.
The quantities $\Vcb$, $\mb$, $\mc$, and nonperturbative 
parameters describing effects of the strong interaction can be 
determined from the measured rates and moments using expansions in 
$1/m_b$ and the strong coupling constant $\alpha_s$ with reliable 
uncertainty estimates.

Various measurements of moments of the hadronic-mass
\cite{Csorna:2004CLEOMoments, 
Aubert:2004BABARMoments, 
Acosta:2005CDFMoments, 
Abdallah:2005DELPHIMoments, 
Schwanda:2007BELLEMassMoments}
and lepton-energy
\cite{Aubert:2004BABARLeptonMoments, 
Abdallah:2005DELPHIMoments,
Urquijo:2006BelleLeptonMoments}
spectra in inclusive semileptonic decays $\semilepXc$ have already been
used for determinations of $\Vcb$, $\mb$, $\mc$, and of four strong-interaction
parameters $\mupi(\mu)$, $\muG(\mu)$, $\rhoD(\mu)$, and $\rhoLS(\mu)$.
The parameters $\mupi(\mu)$ and $\muG(\mu)$ are the expectation values of the kinetic and
chromomagnetic dimension-five operators, respectively, and appear at ${\cal O}(1/\mb^2)$ 
in the expansion. The parameters $\rhoD(\mu)$ and $\rhoLS(\mu)$ are the  
expectation values of the Darwin and spin-orbit dimension-six operators, respectively,
and appear at ${\cal O}(1/\mb^3)$ in the expansion \cite{Benson:2003GammaKineticScheme}.
Here, $\mu$ denotes the Wilson factorization scale that separates
effects from long- and short-distance dynamics.

Combined fits to the $\semilepXc$ moments and moments of the photon-energy spectrum
in $\BtoXsGamma$ decays \cite{Chen:2001CLEOXsGammaMoments, 
Koppenburg:2004BELLEXsGammaInclusive, Aubert:2005BABARXsGammaExclusive,
Aubert:2006XsGammaInclusive, Aubert:2007XsGammaBReco}
in the context of Heavy Quark Expansions (HQE) lead to
$\Vcb = (41.96 \pm 0.23 \pm 0.69) \times 10^{-3}$ and  
$\mb = (4.590 \pm 0.025 \pm 0.030) \gevcc$
in the kinetic-mass scheme \cite{Buchmuller:2005globalhqefit} and
$\Vcb = (41.78 \pm 0.30 \pm 0.08) \times 10^{-3}$ and 
$\mb = (4.701 \pm 0.030) \gevcc$
in the 1S scheme \cite{HFAG:2007LPupdate1SFit}.
The Belle Collaboration has presented similar results in 
\cite{Schwanda:2008BellePhotonMomentsAndFit}.

While lepton-energy moments are known with good accuracy, the precision 
of the hadronic-mass and photon-energy moments is limited by statistics. 
Therefore, we present a new measurement of the hadronic-mass moments $\mxmom{k}$ with
$k=1,\ldots,6$ based on a larger dataset than previously used \cite{Aubert:2004BABARMoments}.
We also present the first measurement of the combined hadronic mass-and-energy moments
$\moment{\nxn}$ with $k=2,4,6$  as proposed by Gambino and Uraltsev 
\cite{Gambino:2004MomentsKineticScheme}. 
The combined moments $\moment{\nxn}$ use the mass $\mx$ and the
energy $\Ex$ of the $\Xc$ system in the $\B$ meson rest frame of $\semilepXc$ decays,
\begin{equation}\label{eq:nxDef}
    n_X^2 = m_X^2 c^4 - 2 \tilde{\Lambda} E_X + \tilde{\Lambda}^2,
\end{equation}
with  a constant $\tilde\Lambda$, here fixed to be 0.65\,\gev as proposed 
in \cite{Gambino:2004MomentsKineticScheme}. They are expected to allow a more 
reliable extraction of the higher-order nonperturbative HQE
parameters and thus to increase the precision on the extraction of $\Vcb$ and the
quark masses $\mb$ and $\mc$. All moments are determined
for different values of the minimum energy of the charged lepton.

We update our previous measurement of lepton-energy moments \cite{Aubert:2004BABARLeptonMoments} 
using branching fraction measurements for background decays in \cite{Yao:2007pdgupdate} 
and improving the evaluation of systematic uncertainties.

Finally, we perform a combined fit to the hadronic-mass moments, moments of
the lepton-energy spectrum, and moments of the photon-energy spectrum in
decays $\BtoXsGamma$. The fit determines $\Vcb$, the quark masses $\mb$ and $\mc$,
the total semileptonic branching fraction $\brf(\semilepXc)$, and the dominant
nonperturbative HQE parameters $\mupi$, $\muG$, $\rhoD$, and $\rhoLS$.
An alternative fit to the moments of $\nxn$, of the lepton-energy, and of the photon
energy in $\BtoXsGamma$, leads to essentially the same results.

\section{\babar\ Detector and Datasets}
\label{sec:detector}

The work is based on data collected with the \babar\ experiment 
\cite{Aubert:2001detector} at the \pep2 asymmetric-energy \epem storage 
rings \cite{Slac:1993pep2} at the SLAC National Accelerator Laboratory.

The $\babar$ tracking system used for charged particle and vertex
reconstruction has two main components: a silicon vertex tracker
(SVT) and a drift chamber (DCH), both operating within a 1.5-T
magnetic field of a superconducting solenoid.
The transverse momentum resolution is $0.47\,\%$ at $1\gevc$.
Photons are identified in an electromagnetic calorimeter (EMC)
surrounding a detector of internally reflected Cherenkov light
(DIRC), which associates Cherenkov photons with tracks for particle
identification (PID). The energy of photons is measured with 
a resolution of $3\,\%$ at $1\gev$. Muon candidates are identified with the
use of the instrumented flux return (IFR) of the solenoid.
The tracking system, EMC, and IFR cover the full azimuthal range 
and the polar-angle range $0.3 < \theta < 2.7\rad$ in the laboratory frame,
corresponding to a coverage of approximately $90\%$
in the center-of-mass (c.m.) frame, where $\theta$ is the polar angle 
with respect to the electron direction. 
The DIRC fiducial volume corresponds to a c.m.~frame coverage of about $84\%$.

The data sample for the hadronic moments measurements consists of about $210\,\invfb$,
corresponding to $232 \times 10^{6}$  decays $\FourS \to \BB$.
Our previous measurement of the lepton-energy moments, which is
updated in this paper, was based on a data sample of about
$51 \times 10^{6}$ $\FourS \to \BB$ decays. This corresponds to an integrated 
luminosity of $47\,\invfb$ on the $\FourS$ resonance. In addition, 
about $9\,\invfb$ of data recorded at an energy $40\mev$ below the resonance 
(off-resonance) was used in the lepton-energy moments measurement 
for the subtraction of background not originating from the $\FourS$.

We use Monte Carlo (MC) simulated events to determine background distributions and 
to correct for detector acceptance and resolution effects. Simulated $\B$-meson 
decays are generated using \evtgen \cite{Lange:2001EvtGen}. The simulation 
of the $\babar$ detector is realized with $\geantf$ \cite{Agostinelli:2002Geant4}
and final state radiation (FSR) is modeled using the $\photos$ code \cite{Richter-Was:1992PHOTOS}.

In the simulation of semileptonic decays $\semilepXc$ we use 
the branching fractions listed in Table \ref{tab:detector:signalbf}.
For the dominant decay $\semilepDstar$ we use a parameterization of form factors,
based on heavy quark effective theory (HQET) 
\cite{Bigi:HQET,Caprini:FormFactors,Grinstein:HQET}.
Its  differential rate is described by three helicity amplitudes 
which are expressed by the three parameters $\rho^{2}$, $R_1$, and $R_2$.
We choose the following values measured in \cite{Duboscq:1996FormFactor}: 
$R_1 = 1.18 \pm 0.30 \pm 0.12$, $R_2 = 0.71 \pm 0.22 \pm 0.07$, and 
$\rho^{2} = 0.91 \pm 0.15 \pm 0.06$. The quoted errors reflect statistical and 
systematic uncertainties.
For decays $\semilepD$ and for decays to the higher-mass states
$\D_1$, $\D_1'$, $\D_0^*$, and $\D_2^*$ we use the 
ISGW2 model \cite{Scora:1995FormFactor}.
For the decays $\semilepNreso$, we use the prescription by 
Goity and Roberts \cite{Goity:1995SoftPion}.

\section{Reconstruction of Semileptonic Decays for the Measurement of Hadronic Moments}
\label{sec:semilep_decays}

The event selection and reconstruction for the hadronic-mass moments
$\moment{m_X^k}$ and the combined mass-and-energy moments
$\moment{n_X^k}$ are almost identical. As described in the
corresponding sections \ref{sec:hadronic_mass_moments} and \ref{sec:mixed_moments}, 
the only differences regard the requirements needed to ensure
a good resolution in the observables of interest.

The analysis uses $\FourS \to \BB$ events in which one of the $\B$ mesons 
decays to hadrons and is fully reconstructed ($\Breco$),
and the semileptonic decay of the recoiling $\Bb$ meson ($\Brecoil$)
is identified by the presence of an electron or muon. While this approach
results in a low overall event selection efficiency of only a few per mille,
it allows for the determination of momentum, charge, and flavor of the $\B$ mesons.

\subsection{Selection of Hadronic $\B$-Meson Decays}
\label{sec:hadronic_mass_moments:breco}

To obtain a large sample of $\Breco$-mesons,  many exclusive hadronic decays 
$\Breco \rightarrow \DbarorDbarstar Y^{\pm}$ are reconstructed \cite{Aubert:2003VubRecoil}. 
The hadronic system $Y^{\pm}$ consists of hadrons with a total charge of $\pm 1$.
It is composed of $\text{n}_{\pi} \pipm$, $\text{n}_{\kaon}\Kpm$, $\text{n}_{\KS} \KS$,
and $\text{n}_{\piz} \piz$ with $\text{n}_{\pi} + \text{n}_{\kaon} \leq 5$, 
$\text{n}_{\KS} \leq 2$, and $\text{n}_{\piz} \leq 2$, respectively. In total
$1097$ hadronic decay modes are reconstructed.

The kinematic consistency of the $\Breco$ candidates is checked with two 
variables, the beam-energy-substituted mass $\mes = \sqrt{s/4 - \vec{p}^{\,2}_B}$ 
and the energy difference $\Delta E = E_B - \sqrt{s}/2$. Here $\sqrt{s}$ is the total
energy in the c.m.\ frame, and $\vec{p}_B$ and $E_B$ denote the c.m.\ momentum 
and c.m.\ energy of the $\Breco$ candidate, respectively. 
The mass $\mes$ is measured with a resolution of $2.5\mevcc$, essentially independent 
of the $\Breco$ channel. We require $\Delta E = 0$ within 
three standard deviations,  where one standard deviation ranges between $10$ 
and $30\mev$ depending on the number of charged and neutral hadrons 
in the $\Breco$ candidate. For each of the reconstructed hadronic modes the purity
is estimated as the fraction of signal decays with $\mes > 5.27 \gevcc$. We restrict 
the selection to hadronic modes with purities of at least $28\%$ resulting in a selected 
$\Breco$ sample with an overal purity of $60\%$. On average we reconstruct $\Breco$ candidates
with an efficiency of about $0.4\%$.

\begin{table}[t]
\begin{center}
\caption{Summary of branching fractions of semileptonic decays $\semilepXc$ 
         used in MC simulations for neutral ($\BR_{\Bz}$) and 
         and charged ($\BR_{\Bpm}$) $\B$-meson decays. 
         The values are taken from \cite{Yao:2007pdgupdate, Barberio:2006hfag, 
         Aubert:2007RelSemBF, Aubert:2007JensSemilepDzero, Aubert:2008HaukeDstarstar}.
         Isospin symmetry is assumed to calculate the individual decay rates for $\Bz$ and $\Bpm$ mesons
         from their averaged measured branching fractions, taking into account the lifetime ratio
         $1.071 \pm 0.009$ \cite{Yao:2007pdgupdate}.
         The sum of the exclusive decays is constrained to equal the total inclusive branching fractions for
         $\semilepXc$ decays, $\brf(\semilepBpXc) = 10.89 \pm 0.16$ and $\brf(\semilepBzXc) = 10.15 \pm 0.16$
         \cite{Barberio:2006hfag, Aubert:2005BABRVubEndpoint}.\\
         }

\begin{tabular}{lcc}
\hline \hline
Semileptonic Decay     & $\BR_{\Bz}$ $[\%]$         & $\BR_{\Bpm}$ $[\%]$\\
\hline
\semilepD         &2.13 $\pm$ 0.14         &2.30 $\pm$ 0.16  \\
\semilepDstar     &5.53 $\pm$ 0.25         &5.95 $\pm$ 0.24  \\
\hline 
\semilepDone      &0.50 $\pm$ 0.08         &0.54 $\pm$ 0.06   \\
\semilepDtwostar  &0.39 $\pm$ 0.07         &0.42 $\pm$ 0.08   \\
\semilepDzstar    &0.43 $\pm$ 0.09         &0.45 $\pm$ 0.09   \\
\semilepDoneprime &0.40 $\pm$ 0.20         &0.45 $\pm$ 0.20   \\
\hline 
$\Bbar \rightarrow \Dz \pi \ell \nu$ 
                  &0.40 $\pm$ 0.12         &0.20 $\pm$ 0.06   \\
$\Bbar \rightarrow \Dpm \pi\ell\nu$ 
                  &0.19 $\pm$ 0.06         &0.40 $\pm$ 0.12   \\
$\Bbar \rightarrow \Dstarz \pi\ell\nu$ 
                  &0.12 $\pm$ 0.04         &0.06 $\pm$ 0.02   \\
$\Bbar \rightarrow \Dstarpm \pi\ell\nu$ 
                  &0.06 $\pm$ 0.04         &0.12 $\pm$ 0.04   \\
\hline \hline
\end{tabular} 

\label{tab:detector:signalbf}
\end{center}
\end{table}

\subsection{Selection of Semileptonic Decays}
\label{sec:hadronic_mass_moments:selection}

Semileptonic  decays are identified by the presence of one and 
only one electron or muon above a minimum momentum $\plmin$
measured in the rest frame of the $B$ meson.
If not stated otherwise, $\plep$ will denote in the following 
the lepton momentum measured in the $\B$-meson rest frame.
Electrons are identified by combining information
from the EMC, the DCH, and the DIRC. They are required to
have a lab-frame momentum of $p > 0.8\gevc$ and a polar angle
in the range $0.41 < \theta < 2.54\rad$. In this range, electrons are selected 
with $94\%$ average efficiency and a hadron misidentification 
rate of the order of $0.1\%$. 
Muon identification is mainly based on information obtained from 
the IFR. Muons are identified with an efficiency ranging between $60\%$ 
for momenta $p = 1\gevc$ in the laboratory frame and $75\%$ for 
momenta $p > 2\gevc$. The misidentification rate ranges
between $1\%$ for kaons and protons and $3\%$ for pions. 
Efficiencies and misidentification rates are estimated
from control samples of electrons, muons, pions, and kaons.
We impose the condition $Q_b Q_{\ell} < 0$, where $Q_{\ell}$ is 
the charge of the lepton and $Q_b$ is the charge of the $b$ quark 
of the $\Breco$. This condition is fulfilled for primary leptons originating
directly from the $\B$ decay, except for $\BzBzb$ events in which 
flavor mixing has occurred.
We require the total observed charge of the event to be 
$|Q_{\rm tot}|= |Q_{\rm \Breco} + Q_{\rm \Brecoil}| \leq 1$,
allowing for a charge imbalance in events with low momentum tracks 
or photon conversions. In cases where only one charged track is present 
in the reconstructed $\Xc$ system,  the total charge in the event is 
required to be zero.

\subsection{Reconstruction of the Hadronic System}
\label{sec:hadmomrecobase_reconstr}

The hadronic system $\Xc$ in the decay \semilepXc\ is reconstructed from charged
tracks  and energy deposits in the calorimeter that are not associated to
the $\Breco$ or the charged lepton.
We ignore tracks and energy deposits in the calorimeter which are
compatible with the hypothesis of being reconstruction artifacts,
low-energy beam-generated photons or calorimeter deposits originating
from hadronic showers.
Each track is assigned a specific
particle type, either $\porantip$, $\Kpm$, or $\pipm$, based on combined information
from the different $\babar$ subdetectors. Few events containing single protons are
kept in the selection but removed later on in the background removal procedure.
The four-momentum $P_{\Xc}$ of the 
reconstructed hadronic system is obtained from the four-momenta of the
reconstructed tracks $P_{i,\mathrm{trk}}$ for the given mass assignment, 
and photons $P_{i,\gamma}$ by
$P_{\Xc} = \sum_{i=1}^{N_{\mathrm{trk}}} P_{i,\mathrm{trk}} + \sum_{i=1}^{N_{\gamma}} P_{i,\gamma}$.
The hadronic mass is defined by $\mx^{2} = P_{\Xc}^{2}$.

The four-momentum of the unmeasured neutrino is calculated from the
missing four-momentum  $\Pmissfourmom = P_{\FourS} - P_{\Breco} - P_{\Xc} - P_\ell$.
Here, all four-momenta are measured in the laboratory frame.
To ensure a well reconstructed hadronic system, we impose criteria on
the missing energy, $\emiss > 0.5 \gev$, the missing momentum, $\pmiss > 0.5 \gevc$,
and the difference of both quantities, $\epmissabs < 0.5 \gev$.

We perform a kinematic fit exploiting the fact that $\B$ mesons are
produced in a well-defined initial state $\epem \rightarrow \FourS \rightarrow \B \Bbar$, 
to improve the resolution and reduce the bias on the reconstructed quantities.
The fit imposes four-momentum conservation, the equality of the masses of the two $\B$
mesons, and constrains the mass of the neutrino, $\Pmissfourmom^{2} = 0$. 
The resulting (original) average resolutions on the measurement of $\mx$
and $\nx$ are $0.355 \gevcc$ ($0.425 \gevcc$) and $\resolNx$ ($\resolNxbefore$), respectively.
The average biases of $\mx$ and $\nx$ after (before) the kinematic fit 
are found to be $-0.096 \gevcc$ ($-0.254 \gevcc$)
and $\biasNx$ ($\biasNxbefore$), respectively.

The background is composed of $\epem \rightarrow \q\qbar\, (\q = u,d,s,c)$ events
(continuum background), $\FourS \rightarrow \BpBm$  or $\BzBzb$ decays  in which
the $\Breco$ candidate is mistakenly  reconstructed from particles coming from both
$\B$ mesons in the event (combinatorial background), and non-signal decays
of the recoiling $\Brecoil$ meson (residual background).
Signal events where the hadronic system is not fully
reconstructed are not considered as an additional source of
background. The effect of missing tracks and photons on the resolution
of the kinematical quantities of interest is taken into account 
by applying the correction procedures described below.

\begin{figure}[t]
   \begin{center}
   \includegraphics{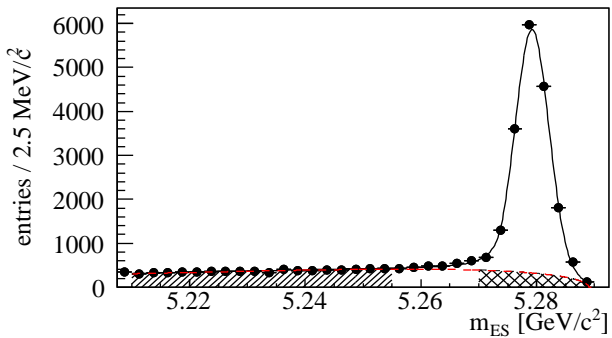}
   \end{center}
   \caption{The $\mes$ spectrum of $\Breco$ decays accompanied by a lepton with $\plgeq{0.8}$.
            The fit functions for the sum of signal and background (solid line)
            and the background (red dashed line) are overlaid.
            The crossed area shows the predicted background under the $\Breco$ signal. 
            The background control region in the $\mes$ sideband is indicated by the hatched area.
            }
    \label{fig:mesFits}
\end{figure}

To quantify the amount of continuum and combinatorial background in the $\mes$ signal region
we perform a fit to the $\mes$ distribution of the $\Breco$ candidates.
We parameterize the background using an empirical threshold
function \cite{Albrecht:1987argusFunction},
\begin{equation}
    \frac{ \text{d}N }{\text{d} \mes } \propto \mes \sqrt{ 1 - x^{2} }
           e^{ -\chi \left( 1 - x^{2} \right) },
\end{equation}
where $x = \mes / \mesmax$, $\mesmax = 5.289\gevcc$ is the kinematic endpoint 
approximated by the mean c.m. energy, and $\chi$ is a free parameter defining 
the curvature of the function. The signal is parameterized with a modified 
Gaussian function \cite{Skwarnicki:1986cbFunction} peaked at the $\B$-meson mass 
and corrected for radiation losses. The fit is performed separately for several 
bins in $\mx$ and $\nx$ to account for changing background contributions. 
Figure \ref{fig:mesFits} shows the $\mes$ distribution for $\plgeq{0.8}$ together 
with the fitted signal and background contributions.
The shape of the continuum and combinatorial background as function of 
$\mx$ and $\nx$ is determined in a signal-free region of the $\mes$ sideband, 
$5.210 \leq \mes \leq 5.255 \gevcc$. Its overall size in the 
$\mes$ signal region, $\mes > 5.27\gevcc$, is determined
by rescaling with the relative background contributions in the signal 
and sideband regions as determined by the fit. Signal and sideband region are separated 
by $15\mevcc$ to avoid the leakage of signal events into the sideband region.

Residual background is estimated from MC simulations. It is composed of
charmless semileptonic decays $\semilepXu$, hadrons misidentified
as leptons, secondary leptons from semileptonic decays of $\DorDstar$, $\Ds$ mesons 
or $\tau$ either in $\BzBzb$ mixed events or produced in $\b \rightarrow \c \cbar \s$ 
transitions, as well as leptons from decays of $\jpsi$ and $\psitwos$. 
The branching fractions of the individual simulated background decays are scaled to
agree with measurements \cite{BABAR:BtoDUppperVertex, Adam:2006dsemilep, 
Yao:2007pdgupdate, Barberio:2006hfag}. The overall simulated background spectrum is 
normalized to the number of $\Breco$ events in data. 
We verify the normalization and the shape using an independent data control 
sample with inverted lepton charge correlation, $Q_b Q_{\ell} > 0$.

\section{Hadronic-Mass Moments}
\label{sec:hadronic_mass_moments}

We present measurements of the moments $\mxmom{k}$, with $k=1,\ldots6$,
of the hadronic-mass distribution in semileptonic $\B$-meson decays $\semilepXc$.
The moments are measured as functions of the lower limit on the lepton 
momentum $\plmin$ between $0.8\gevc$ and $1.9\gevc$, calculated in the 
rest frame of the $\B$ meson.

\subsection{Selected Event Sample}

\begin{figure}
   \begin{center}
   \includegraphics{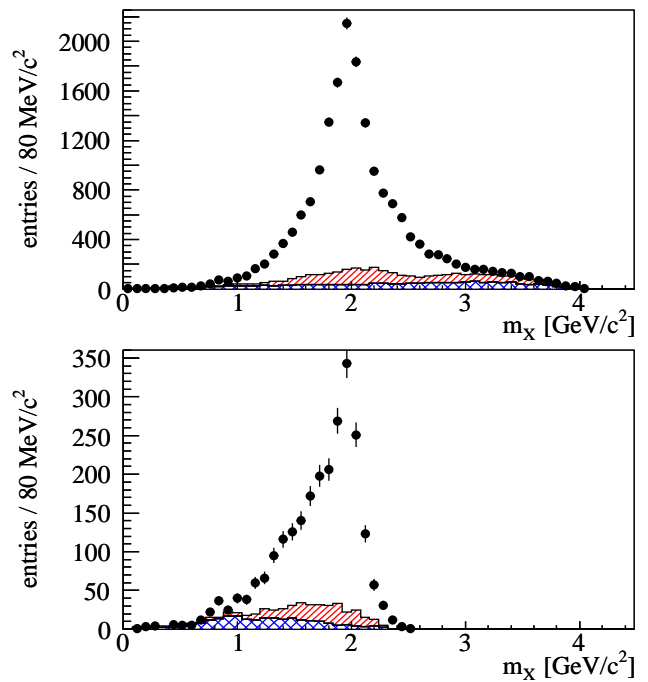}
   \end{center}
   \caption{Hadronic-mass spectra after the kinematic fit 
            for lepton momenta $\plgeq{0.8}$ (top) and $\plgeq{1.9}$ (bottom)
            together with distributions of combinatorial background
            and background from non-$\BB$ decays (red, hatched area) 
            as well as residual background (blue, crossed area). 
            The two background histograms are plotted on top of each other.
           }
    \label{fig:mass_spectra}
\end{figure}

We find $19,212$ events with $\plgeq{0.8}$, composed of
$15,085 \pm 146$ signal events above a combinatorial and continuum 
background of $2,429 \pm 43$ events and residual background 
of $1,696 \pm 19$ events. 
Signal decays amount to $79\%$ of the selected event sample.
For $\plgeq{1.9}$, we find in total $2,527$ events composed of
$2,006 \pm 53$ signal events above a background of $271 \pm 17$ and $248 \pm 7$ 
combinatorial/continuum and residual events, respectively.
Figure \ref{fig:mass_spectra} shows the $\mx$ distributions 
after the kinematic fit together with the extracted background shapes for 
$\plgeq{0.8}$ and $\plgeq{1.9}$.

\subsection{Extraction of Moments}
\label{sec:hadronic_mass_moments:extraction}

To extract unbiased moments $\mxmom{k}$, we apply corrections to account 
for effects that distort the measured $\mx$ distribution.
Contributing effects are the limited acceptance and resolution of the $\babar$
detector resulting in unmeasured particles and in misreconstructed energies and momenta of
particles. In addition, there are contributions from measured particles not belonging to 
the hadronic system, especially photons originating from FSR 
of the primary leptons. These photons are included in the measured $\Xc$
system and thus lead to a modified value of its mass; they also lower the 
momentum of the primary lepton. Both effects are included in our correction procedure.

We correct the kinematically-fitted value of $\mx^{k}$ of each event by applying
correction factors on an event-by-event basis using the
observed linear relationship between the moments of the measured mass $\mxmomreco{k}$
and the moments of the true mass $\mxmomtruecut{k}$ in MC spectra. The correction factors
are determined from MC simulations by calculating moments $\mxmomreco{k}$ and
$\mxmomtruecut{k}$ in several bins of the true mass $\mxtrue$ and fitting the
observed dependence with a linear function, referred to as calibration function
in the following.

\begin{figure*}
   \begin{center}
   \includegraphics{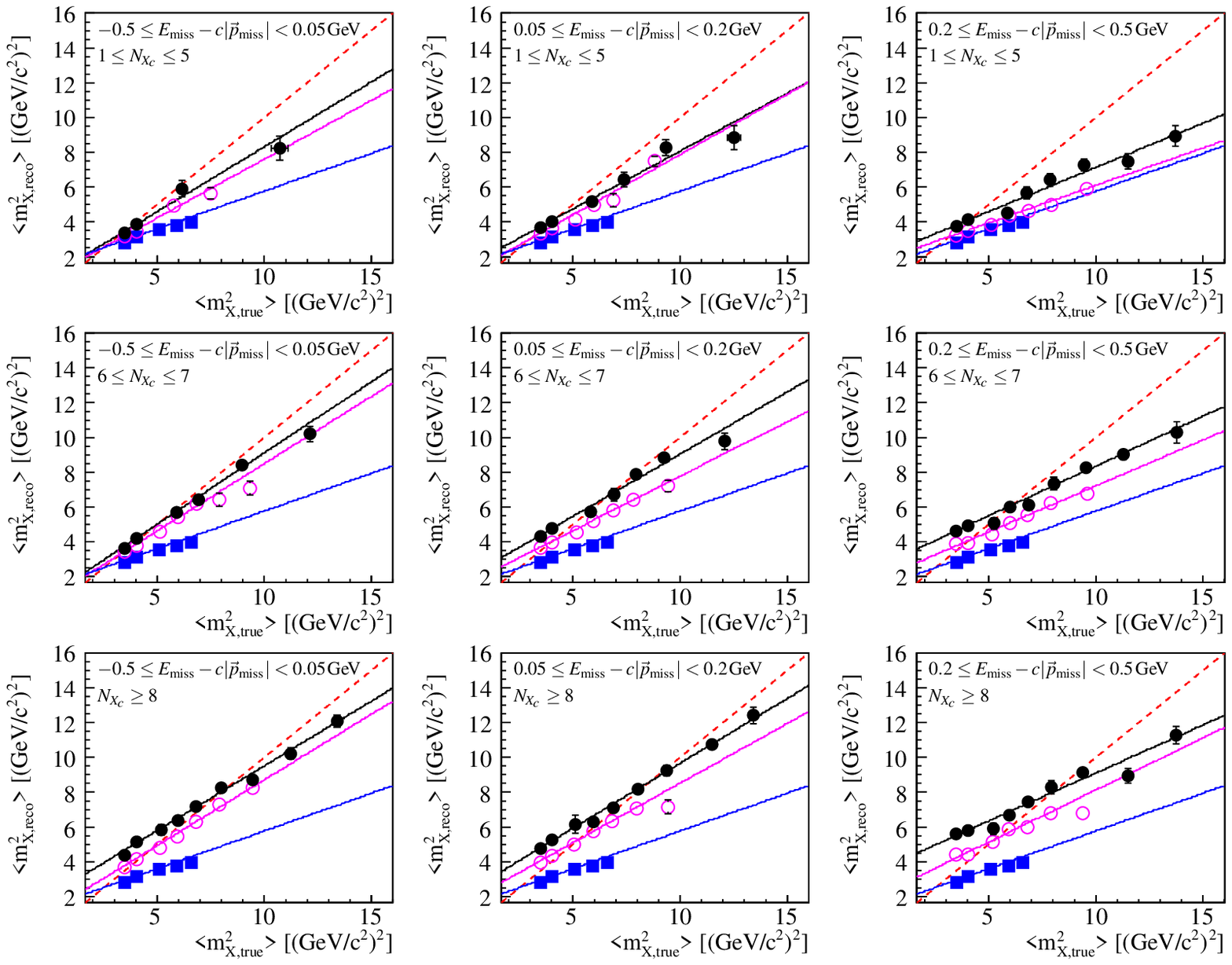}
   \end{center}
   \caption{Examples of calibration functions for $\mxmom{2}$ in bins of $\MultX$, $\epmiss$ and
            $\plep$. Shown are the extracted moments $\mxmomreco{2}$ versus the true
            moments $\mxmomtruecut{2}$ for $\plbin{0.8}{0.9}$ (\textcolor{black}{$\bullet$}),
            $\plbin{1.4}{1.5}$ (\textcolor{magenta}{$\circ$}), and
            $\plgeq{1.9}$ (\textcolor{blue}{$\blacksquare$}). The results of fits of linear
            functions are overlaid as solid lines. A reference line with
            $\mxmomreco{2} = \mxmomtruecut{2}$ is superimposed (dashed line).
            The calibration function for $\plgeq{1.9}$ is constructed independent of
            $\MultX$ and $\epmiss$. It is plotted in each of the bins for comparison only.
           }
    \label{fig:calib_mx2}
\end{figure*}

We find that the bias of the measured moments $\mxmomreco{k}$ is 
not constant over the whole phase space.
Therefore, we derive the calibration functions in three bins of the particle 
multiplicity $\MultX$ in the $\Xc$ system, 
three bins of $\epmiss$, as well as in twelve bins of
$\plep$, each with a width of $100\mevc$. 
Due to the limited number of generated MC events,
the binning in $\MultX$ and $\epmiss$ is not used for $\plmin \geq 1.7\gevc$.
Overall we construct $84$ calibration functions for each order of moments.
The obtained calibration functions allow a consistent extraction of moments
for events containing an electron or a muon.
Figure \ref{fig:calib_mx2} shows examples of calibration functions for the moment $\mxmom{2}$
in three bins of $\plep$ as well as in nine bins of $\epmiss$ and $\MultX$.

For each data event $i$, the corrected mass $\mxcalibi^{k}$ is calculated by 
inverting the linear function,
\begin{equation}
    \mxcalibi^{k} = \frac{\mxrecoi^{k} - A(\epmiss, \MultX, k, \plep)}
                         {B(\epmiss, \MultX, k, \plep)},
\end{equation}
where $A$ is the offset and $B$ is the slope of the calibration function. 
Background contributions are removed by applying a weight factor
$\wi$ to each corrected hadronic mass $\mxcalibi^{k}$, where the weight is the
expected fraction of signal events in the corresponding region of the 
$\mxreco$ spectrum in Fig.~\ref{fig:mass_spectra}.
The expression used to calculate the moments is the following:
\begin{equation}
   \mxmom{k} = \frac{\sum\limits_{i=1}^{N_{\mathit{ev}}} \wi(\mx) \, \mxcalibi^{k}}
                    {\sum\limits_{i}^{N_{\mathit{ev}}} \wi}
               \times \Ccalib(\plep, k) \times \Ctrue(\plep, k),
\end{equation}
with $N_{\mathit{ev}}$ the total number of selected events.
The factors $\Ccalib$ and $\Ctrue$ depend on the order $k$ and
the minimum lepton momentum $\plmin$ of the measured moment.
They are determined in MC simulations and correct
for the residual small biases observed after the calibration.
The factors $\Ccalib$ account for the bias of the applied correction method and
are typically ranging between 1.01 and 1.06 for $k=1 \ldots 5$. 
Larger bias corrections $\Ccalib$ are observed for $\mxmom{6}$ 
ranging between the extremes $0.902$ and $1.109$.
The residual bias-correction factor $\Ctrue$
accounts for differences in selection efficiencies for different hadronic
final states and FSR that is included
in the measured hadron mass and distorts the measurement of the lepton's momentum.
Our correction procedure results in moments which are within systematic uncertainties
free of photon radiation.
The correction $\Ctrue$ is estimated in MC simulations 
and typically ranges between $0.994$ and $1.007$.
For the moments $\mxmom{5}$ and $\mxmom{6}$, slightly higher correction factors
are determined, ranging between $0.990$ and $1.014$ for $\mxmom{5}$
and $0.986$ and $1.024$ for $\mxmom{6}$.

\begin{figure}
   \begin{center}
   \includegraphics{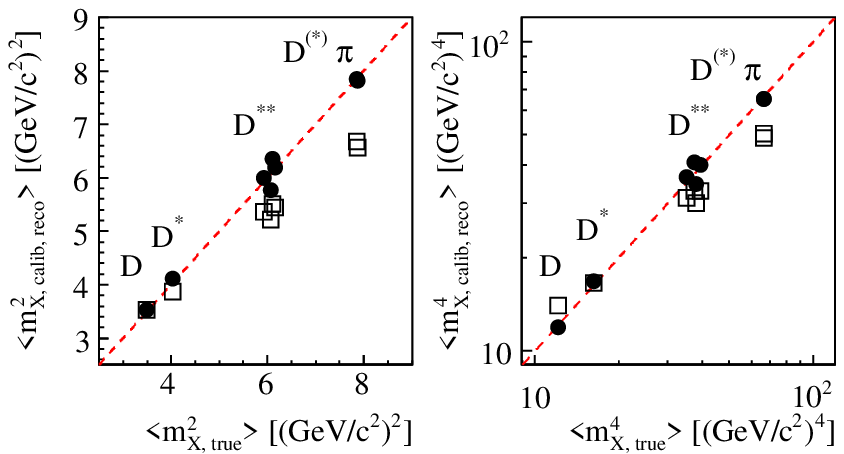}
   \end{center}
   \caption{Calibrated ($\bullet$) and uncorrected ($\Box$) moments
            $\mxmom{2}$ (left) and $\mxmom{4}$ (right)
            of individual hadronic modes for lepton momenta $\plgeq{0.8}$.
            A reference line with $\mxmomcalib{} = \mxmomtruecut{}$ is superimposed.
           }
    \label{fig:massMoments_exclusiveModes}
\end{figure}

This correction procedure is verified on a MC sample by applying the calibration to
measured hadron masses of individual semileptonic decays, $\semilepD$,
$\semilepDstar$, four resonant decays $\semilepDstarstar$,
and two nonresonant decays $\semilepNreso$.
Figure \ref{fig:massMoments_exclusiveModes} shows the corrected
moments $\mxmom{2}$ and $\mxmom{4}$ as functions of the true moments
for minimum lepton momenta $\plgeq{0.8}$.
The dashed line corresponds to ${\protect \mxmomcalib{k}} = {\protect \mxmomtruecut{k}}$.
The calibration reproduces the true moments over the full mass range.

\subsection{Systematic Uncertainties and Tests}
\label{sec:hadronic_mass_moments:systematics}

The main systematic uncertainties are associated with the modeling of hadronic final states
in semileptonic $\B$-meson decays, the bias of the calibration method, the determination
of residual background contributions, the modeling of track and photon selection efficiencies,
and the identification of particles. 
The total systematic uncertainty is estimated by adding in quadrature 
five contributions, as described below.
Tables \ref{tab:massMomentsSummary_1} and \ref{tab:massMomentsSummary_2} 
list the individual contributions to the systematic errors of the measured 
moments $\mxmom{k}$ with $k = 1 \ldots 6$ for minimum lepton momenta 
ranging from $0.8$ to $1.9 \gevc$.

\subsubsection{MC Statistics}
\label{sec:hadronic_mass_moments:systematics:MCStatistics}

The effect of limited MC statistics on the extracted moments is evaluated using 
parameterized MC experiments. To study the effect on the calibration
curves, the parameters of the fitted first-order polynomials are randomly
varied within their uncertainties including correlations and new sets of moments are extracted.
The overall uncertainty is determined by repeating this procedure 250 times and
taking the r.m.s.~of the distribution of the moments as the systematic uncertainty.

To estimate the effect of limited MC statistics in the residual background determination
a similar method is applied by varying the parameters of the fit to the $\mes$ distribution
within their errors including correlations.

\subsubsection{Simulation-Related Effects}

We correct for differences between data and MC simulation in
the selection efficiencies of charged tracks and photons, 
as well as identification efficiencies and misidentification
rates of various particle types. The corrections are extracted from data and MC
control samples.

The systematic uncertainties of the photon selection and track finding efficiencies 
are determined studying independent control samples.
Their impact on the measured moments has been evaluated 
by randomly excluding neutral or charged candidates from 
the $\Xc$ system with probabilities corresponding to 
the uncertainties of the efficiency extraction methods.
The uncertainty of the photon selection efficiencies is found to be
$1.8\%$ per photon independent of energy, polar angle, and multiplicity. 
The uncertainty in track finding efficiencies consists of two parts.
For each track, we add in quadrature $0.8\%$ systematic uncertainty 
and the statistical uncertainty of the control samples that depend 
on energy and polar angle of the track as well as the multiplicity of tracks 
in the reconstructed event.

The systematic uncertainty on the misidentification of 
$\pipm$ mesons as leptons is found to affect the
overall normalization of the corresponding background spectra by $8\%$.
The influence on the measured moments is estimated by varying the
corresponding background within its uncertainty. The observed variation of
moments is taken as a systematic uncertainty.

The impact of mismodeling FSR simulated with $\photos$ \cite{Richter-Was:1992PHOTOS} is
estimated by calculating moments from data using a set of calibration curves constructed
from events simulated without FSR photons. The theoretical uncertainty associated with
the calculations included in $\photos$ is conservativley assumed to be of the order of $20\%$.
The systematic uncertainty connected to the mismodeling of FSR photons is therefore
estimated to be $20\%$ of the observed difference between the nominal moments and
those from the MC simulation without FSR photons.

A significant fraction of the low-energy photons detected in the calorimeter are beam related.
We check the impact of low-energy photons by removing
EMC signals with energies below $100 \mev$ from the reconstructed hadronic
system. The effect on the measured moments is found to be negligible.

The stability of the result under variation
of the selection criteria on $\epmiss$ is tested by varying the applied cut
between $\epmissabs < 0.2\gev$ and $\epmissabs < 1.4\gev$. For all measured 
moments, the observed variation is well covered by other known systematic
detector and MC simulation effects. Therefore, no contribution is added to the systematic
uncertainty.

\subsubsection{Extraction Method}

The systematic uncertainty of the calibration bias correction $\Ccalib$
is estimated to be $(\Ccalib - 1)/2$.

\subsubsection{Background Determination}

The branching fractions of background decays in the MC simulation are scaled to
agree with the current measurements \cite{BABAR:BtoDUppperVertex, Adam:2006dsemilep, 
Yao:2007pdgupdate, Barberio:2006hfag}. The associated systematic
uncertainty is estimated by varying these branching fractions within their 
uncertainties. At low $\plmin$, most of the studied background channels contribute to 
the systematic uncertainty equally, while at high $\plmin$, the systematic uncertainty 
is dominated by background from decays $\semilepXu$.
Contributions from $\jpsi$ and $\psitwos$ decays are found to be negligible.

The uncertainty in the combinatorial $\Breco$ background determination is estimated by
varying the lower and upper limits of the sideband region in the $\mes$ distribution
up and down by $2.5\mevcc$. The observed effect on all hadronic-mass moments 
is found to be negligible.

\subsubsection{Modeling of Signal Decays}
\label{sec:hadronic_mass_moments:systematics:Signal}

The uncertainty of the calibration method with respect to the chosen signal model
is estimated by changing the composition of the simulated inclusive hadronic spectrum.
The dependence on the simulation of high mass hadronic final states is estimated
by constructing calibration functions only from MC simulated hadronic events with
hadronic masses $\mxtrue < 2.5 \gevcc$, thereby removing the high mass tail of the simulated
hadronic-mass spectrum. The model dependence of the calibration method is
found to be a small contribution to the total systematic uncertainty.

We estimate the model dependence of the residual bias correction $\Ctrue$
by changing the composition of the inclusive hadronic spectrum, i.e.~omitting
one or more decay modes.

We study the effect of differences between data and 
MC simulation in the multiplicity and $\epmiss$
distributions on the calibration method by changing the binning of the
calibration functions. The observed variation of the results are found to be
covered by the statistical uncertainties of the calibration functions, 
and no contribution is added to the total systematic uncertainty.

\subsubsection{Stability of the Results}
\label{sec:hadronic_mass_moments:systematics:Stability}

The stability of the results is tested by dividing the data
into several independent subsamples: $\Bpm$ and $\Bz$, decays to electrons and
muons, different run periods of roughly equal sample sizes, and two regions
in the $\epmiss$ spectrum, $-0.5 \leq \epmiss < 0 \gev$ and $0 \leq \epmiss < 0.5 \gev$,
characterized by different resolutions of the reconstructed hadronic system.
No significant variations are observed.

\subsection{Results}
\label{sec:hadronic_mass_moments:results}

\begin{figure*}[t]
   \begin{center}
   \includegraphics{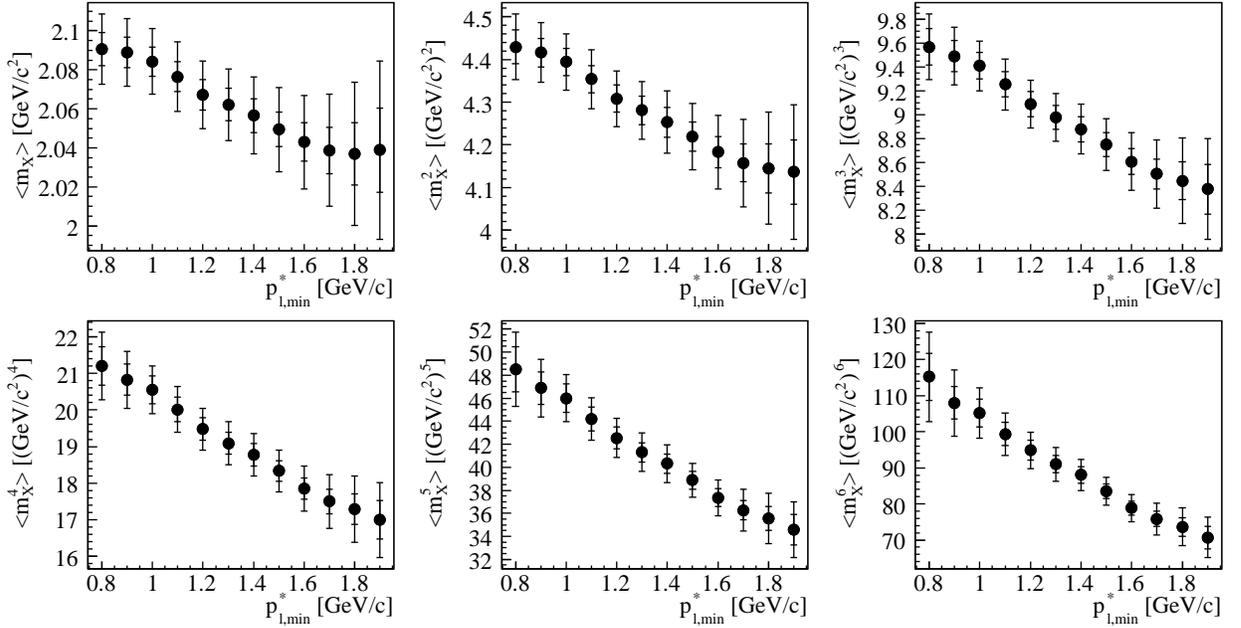}
   \end{center}
   \caption{Radiation-corrected hadronic-mass moments 
            $\mxmom{k}$ with $k = 1 \ldots 6$ for different selection 
           criteria on the minimum lepton momentum $\plmin$. The inner error bars
           correspond to the statistical uncertainties while the full error bars
           correspond to the total uncertainties. 
           The moments, as well as their values for different $\plmin$, 
           are highly correlated.
           }
    \label{fig:massMoments}
\end{figure*}

The measured hadronic-mass moments $\mxmom{k}$ after radiative correction with 
$k = 1 \ldots 6$ as functions of the minimum lepton momentum $\plmin$ are 
shown in Fig.~\ref{fig:massMoments}. All measurements are correlated since 
they share subsets of selected events. Tables \ref{tab:massMomentsSummary_1} 
and \ref{tab:massMomentsSummary_2} summarize the numerical results.
In most cases we find systematic uncertainties that exceed the statistical uncertainty by
a factor of $~2.5$. The correlation matrix for the moments is given in the EPAPS document
\cite{ThisPRD:onlineversion}.

\section{Moments of the Combined Mass-and-Energy Spectrum\label{sec:mixed_moments}}

The measurement of moments of the observable \nx, a combination of the mass and energy of the
inclusive $\Xc$ system, as defined in Eq.~(\ref{eq:nxDef}), is theoretically motivated and is
expected to allow a more reliable extraction of the
higher order HQE parameters \mupi and \rhoD \cite{Gambino:2004MomentsKineticScheme}. 

We present measurements of the moments \moment{\nx}, \moment{\nxfour}, and \moment{\nxsix} for
different minimum lepton momenta between $0.8 \gevc$ and $1.9 \gevc$ in the $\B$-meson rest
frame.

\subsection{Event Selection}

Due to the structure of the variable \nx\ as a difference of two measured values, its 
measured resolution and bias are worse than for the mass moments.
Also, the sensitivity to cuts on $\epmiss$ increases. 
The average resolution of $\nx$ after the kinematic fit for lepton momenta greater than $0.8\gevc$
is measured to be \resolNx\ with a bias of \biasNx. We therefore introduce stronger 
requirements on the reconstruction quality of the event. We tighten the criteria on 
the neutrino observables by requiring $\epmiss$ to be between $\epmisslow$ 
and $\epmissup \gev$.
Due to the stronger requirement, the individual variables
$\emiss$ and $\pmiss$ have less influence on the resolution of the
reconstructed hadronic system. Therefore, the requirements on the missing energy and the missing
momentum in the event are relaxed to $\emiss>\emisslow\,\gev$  and $\pmiss > \pmisslow\,\gevc$,
respectively, as these requirements do not yield significant improvement on the resolution of $\nx$, and do not
increase the ratio of signal to background events.

For $\plgeq{0.8}$ and $1.9\gevc$, there remain \NsigFirst\ and 
\NsigSecond\ signal events, respectively.
Background events make up $\percentBG$ of the final event sample with $\plgeq{0.8}$.
The background is composed of $\percentCombBG$ continuum and 
combinatorial background and $\percentResidualBG$ decays
of the signal $\B$ meson other than the semileptonic decay $\semilepXc$.
Combinatorial and continuum background is removed using the sideband of the $\mes$ distribution, as
described in section \ref{sec:hadmomrecobase_reconstr}. The residual background 
events, containing a correctly reconstructed $\Breco$ meson, are removed using 
MC simulations. The dominant sources are pions misidentified  as muons,
$\semilepXu$ decays, and secondary semileptonic decays of $D$ and $D_s$ mesons.

\begin{figure}[t]
\centering
\includegraphics[width=0.45\textwidth]{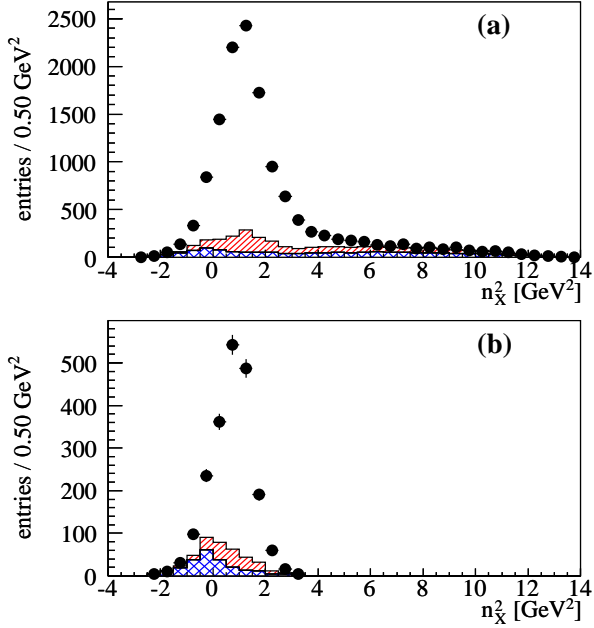}
\caption{Spectra of \nx\ after the kinematic fit together with distributions of
            combinatorial background and background from non-\BB decays
            (red, hatched area) as well as residual background
            (blue, crossed area) for different minimum lepton momenta
            (a) $\plmin=0.8\gevc$ and (b) $\plmin = 1.9\gevc$. 
            The two background histograms are plotted on top of each other.
\label{fig:nxSpectra}} 
\end{figure}

The measured $\nx$ spectra for $\plmin = 0.8\gevc$ and $\plmin = 1.9\gevc$ are 
shown together with the background distributions in Fig.~\ref{fig:nxSpectra}. 

\subsection{Extraction of Moments}

The extraction of unbiased moments $\moment{\nxn}$ from the measured
$\nx$ spectra follows a calibration procedure similar to the one used to extract
the hadronic-mass moments as described in Section \ref{sec:hadronic_mass_moments:extraction}. 
The linear calibration functions
\begin{equation}
n^{k}_{X,\mathrm{calib}}  = \frac{ n^{k}_{X,\mathrm{reco}} -A(\epmiss, \MultX, k, \plep)}{B(\epmiss,
 \MultX, k, \plep)}
\end{equation}
for $k=2,4,6$  are derived from MC samples in three bins of $\epmiss$ and
three bins of the $\Xc$-system multiplicity $\MultX$ for each of the 12 
lepton momentum bins of $100 \mevc$ width. 
Because of differences in events containing electrons and muons,
we also derive separate calibration functions for these two classes of events. 
Overall, we determine $216$ linear calibration functions. 
The calibration again includes the effects of FSR 
photons which not only modify $\mx$ and $\plep$, but also $\Ex$.

\begin{figure}[t]
\centering
\includegraphics[width=0.49\textwidth]{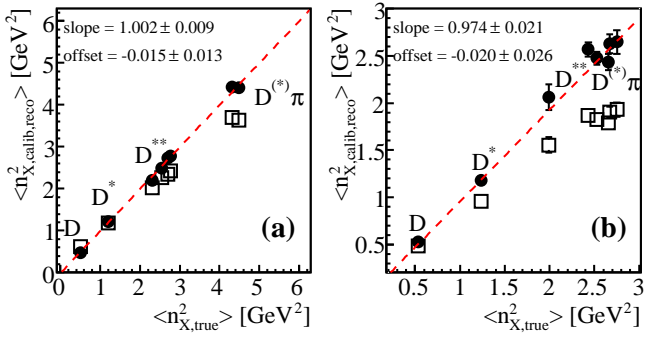}
\caption{Example of the calibration verification procedure for different minimum lepton momenta
         (a) $\plmin = 0.8\gevc$ and (b) $\plmin = 1.7\gevc$. 
         Moments \moment{\nx}\ of exclusive modes on
         simulated events before ($\Box$) and after ($\bullet$)  calibration are
         plotted against the true moments for each mode. The dotted line shows the result 
         of a fit to the calibrated moments, the resulting parameters are given.
         \label{fig:exclModes}}
\end{figure}

We have verified that applying the calibration procedure on MC samples 
of individual exclusive $\semilepXc$ modes allows to reproduce the generated moments,
as shown in Fig.~\ref{fig:exclModes}.
Small biases remaining after calibration are of the order of 1\,\% for \moment{\nx} and of few percent for
\moment{\nxfour}  and \moment{\nxsix}.

\begin{figure*}
   \begin{center}
   \includegraphics[width=0.95\textwidth]{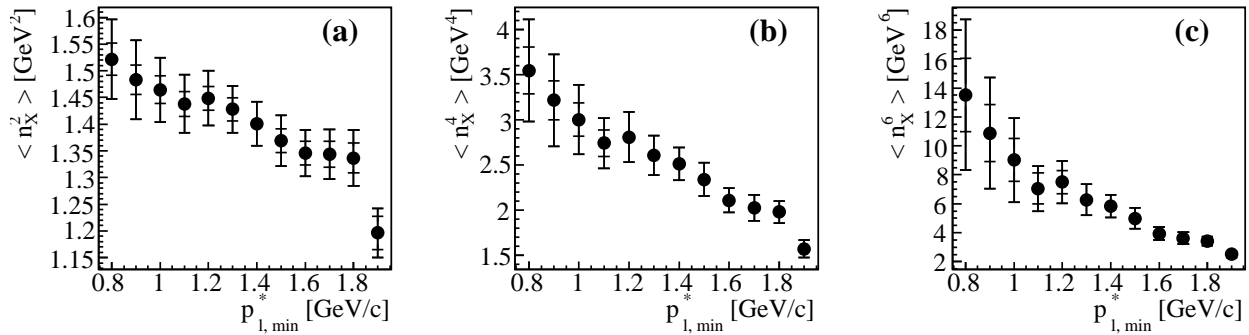}
   \end{center}
   \caption{Radiation-corrected moments (a) \moment{\nx}, 
            (b) \moment{\nxfour}, and (c) \moment{\nxsix} for different cuts on 
            the lepton momentum $\plep$.  The inner error bars correspond to the 
            statistical uncertainties while the full error bars correspond to 
            the total uncertainties. The moments are highly correlated.
            \label{fig:nxMoments}}
\end{figure*}

Background contributions are removed by applying $\nx$-dependent weight factors $w_{i}(\nx)$ on an 
event-by-event basis, leading to the following expression for the determination of the moments:
\begin{equation}
\moment{\nxn} = \frac{\sum\limits_{i=1}^{N_{\mathrm{ev}}} w_{i}(\nx) \,\, {\nxn}_{\mathrm{calib},i} }     
                               {\sum\limits_{i=1}^{N_{\mathrm{ev}}} w_{i}(\nx) } 
         \times 
        \mathcal{C}(\plep, k).
\end{equation}
The bias correction factors $\mathcal{C}(\plep, k)$, depending on the minimum 
lepton momentum and the order of the extracted moments, are determined by MC simulations; 
they combine the two factors $\Ccalib$ and $\Ctrue$ as described in Section 
\ref{sec:hadronic_mass_moments:extraction}.

\subsection{Systematic Uncertainties and Tests}

We consider the same five sources of systematic 
uncertainties as for the mass moments described in 
Sections \ref{sec:hadronic_mass_moments:systematics:MCStatistics} to 
\ref{sec:hadronic_mass_moments:systematics:Signal}: MC statistics, simulation-related effects,
extraction method, background determination, and modeling of signal decays. 
The individual contributions to the systematic error, listed in
Table \ref{tab:NxOrder246}, are estimated following procedures essentially identical  
to those described for the mass moments.

Because of the tighter cut on $\epmiss$, the systematic uncertainty
associated with this criterion is estimated in a different way. We first
keep the lower limit fixed to the nominal value and vary the upper limit
to $0.3 \gevc$ to $0.25\gevc$, $0.4\gevc$, and $0.5\gevc$.
Then we fix the upper limit to its nominal value and vary the lower limit
to $-0.3\gevc$ and $-0.1 \gevc$. The mean of the observed differences in the
measured moments on data is taken as systematic uncertainty.

In the third study, we include the uncertainty from the binning of the 
calibration function in the multiplicity of the $\Xc$-system.
For the choice of the calibration function, 
we randomly increase the measured multiplicity of the $\Xc$ system  by one 
with a probability of $5\%$ corresponding to the observed difference between  
MC and data. The uncertainty in the bias-correction factor $\mathcal{C}(\plep, k)$
is conservatively estimated as half of the applied correction.

Varying the branching fractions of the exclusive signal modes in the 
MC simulation has, in agreement with the mass-moment studies, a very 
small impact on the measured combined moments. Also, no significant variations 
of the results are observed when splitting the data sample into the
same subsamples as for the mass moments.

\subsection{Results}

Figure \ref{fig:nxMoments} shows the results for the moments  \moment{\nx}, 
\moment{\nxfour}, and \moment{\nxsix} 
as a function of the minimum lepton momentum $\plmin$. 
The moments are highly correlated due to the overlapping
data samples. The full numerical results
and the statistical and the estimated systematic uncertainties are given 
in Table \ref{tab:NxOrder246}.
The systematic covariance matrix for the moments of different order and with different cuts on
\plmin\ is built using statistical correlations.
This correlation matrix for the moments is given in the EPAPS document 
\cite{ThisPRD:onlineversion}.

A clear dependence on the minimum lepton momentum is observed 
for all moments, due to the increasing contributions from higher-mass 
final states with decreasing lepton momentum. In most cases we obtain 
systematic uncertainties slightly exceeding the statistical uncertainty.

\section{Moments of the Electron-Energy Spectrum\label{sec:lep_moments}}

\newcommand{\ls}{\ensuremath{e^\pm e^\pm\ }}
\newcommand{\uls}{\ensuremath{e^+ e^-\ }}
\newcommand{\emin}{\ensuremath{E_{0}}\xspace}
\def\Ee    {\ensuremath{E_e}\xspace}

Moments of the electron-energy spectrum for semileptonic decays $\semilepXce$ averaged
over charged and neutral $\B$ mesons have been measured in a data sample
of $51 \times 10^{6}$ $\FourS\to \BB$ decays \cite{Aubert:2004BABARLeptonMoments}. In
the following, we present an overview of this analysis and update the results 
by using more recent measurements \cite{Yao:2007pdgupdate, BABAR:BtoDUppperVertex}
of branching fractions of background processes.

\begin{figure}[b]
\begin{center}
\includegraphics[height=3.2in]{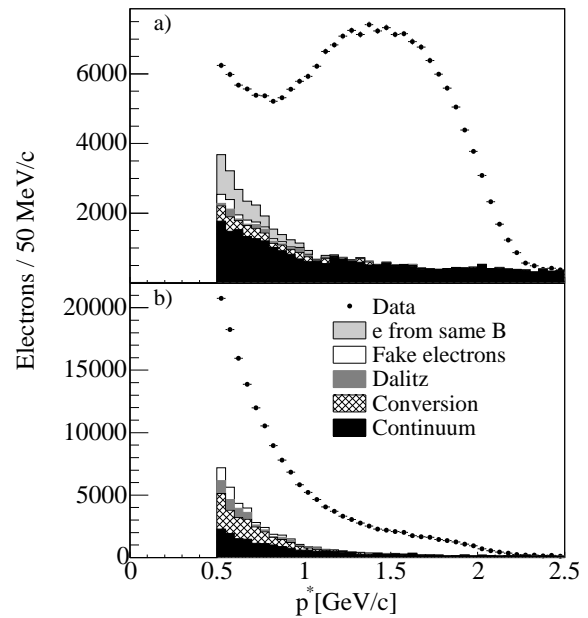} 
\caption{Previously measured momentum spectrum (points) \cite{Aubert:2004BABARLeptonMoments} 
and estimated backgrounds (histograms) for electron candidates in (a) the 
unlike-sign sample, and (b) the like-sign sample.
The background spectra are updated wrt.~the previous publication with more recent
branching-fraction measurements \cite{Yao:2007pdgupdate, BABAR:BtoDUppperVertex}.
}
\label{rawspectra}
\end{center}
\end{figure}

In multi-hadron events as defined in \cite{Aubert:2004BABARLeptonMoments}, \BB events are selected 
by requiring a semileptonic $\B$ decay with an identified electron (\etag), with charge 
$Q(\etag)$ and a momentum $1.4<\plepe<2.3\gevc$, measured in the $\FourS$ rest frame. These
events constitute a tagging sample used as normalization for the branching fraction.
A second electron \esig, for which we require $\plepe>0.5 \gevc$, is assigned either to the
unlike-sign sample if the tagged sample contains an electron with $Q(\etag)=-Q(\esig)$ or to the
like-sign sample if $Q(\etag)=Q(\esig)$.
In events without \BzBzb mixing, primary electrons from semileptonic
\B decays belong to the unlike-sign sample while secondary electrons contribute to the like-sign
sample. Secondary electrons originating from the same \B as the \etag are removed from the 
unlike-sign sample by the requirement
\begin{equation}
\cos\alpha^*   > 1.0 - p_e^* \,\rm{c / \gev}  \ \ \rm{and} \ \   
\cos\alpha^*> -0.2, 
\label{oacut}
\end{equation}
where  $\alpha^*$ is the angle between the two electrons in the 
$\FourS$ rest frame. Corrections for the small residual
background of unlike-sign pairs originating from the same $\B$ fulfilling this requirement 
are described in \cite{Aubert:2004BABARLeptonMoments}. Additional background corrections 
for electrons  from  $\jpsi\to e^+e^-$ decays, continuum events, photon conversions, 
$\pi^0 \to e^+ e^- \gamma$ Dalitz decays, and misidentified hadrons are also described 
in \cite{Aubert:2004BABARLeptonMoments}. Figure \ref{rawspectra} shows the electron-momentum spectra 
together with the contributions of the backgrounds.

Further backgrounds arise from decays of $\tau$ leptons, charmed 
mesons produced in $b \ra c\cbar s$ decays, and $\jpsi$ or 
$\psitwos \to \epem$ decays with only one detected electron. We also need to correct
for cases where the tagged electron does not originate from a semileptonic $B$ decay.
These backgrounds are irreducible. Their contributions to the three 
samples -- single electrons, like-sign, and unlike-sign pairs -- are estimated from 
MC simulations, using the ISGW2 model~\cite{Scora:1995FormFactor} to describe
semileptonic $D$ and $D_s$-meson decays. As an important update to the results in
\cite{Aubert:2004BABARLeptonMoments}, the branching fractions of these backgrounds are 
recalculated to match the recent measurements \cite{Yao:2007pdgupdate}. As in 
\cite{Aubert:2004BABARLeptonMoments}, we calculate $\BR(D_s \to X e \nu) = (7.79 \pm 0.19)\%$
from $\BR(\Dz \to X e \nu)$ and $\BR(D^+ \to X e \nu)$, assuming 
$\Gamma(D_s\to Xe\nu)=\Gamma(D\to Xe\nu)$. Using  
$\BR (B^{0,+} \to D_s^+ X ) = (8.3 \pm 0.8) \%$~\cite{Yao:2007pdgupdate} the branching fraction
of $B^{0,+} \to D_s^+\to e^+$ decays, where the $D_s$ originates from fragmentation 
of the $W$ boson, is $(0.65 \pm 0.06)\%$. Using the inclusive 
branching fraction measurement of $\B^{0,+} \to \D^{0,+} X$ decays reported in
\cite{BABAR:BtoDUppperVertex}, we arrive at $\BR(B^{0,+} \to D^{0,+} \to e^+) = (0.93 \pm 0.11)\%$.
To estimate the contribution of electrons from $\tau$ decays,
we consider the cascades $B \to \tau \to e$ and 
$B \to D_s \to \tau \to e$, with branching fractions taken from~\cite{Yao:2007pdgupdate}.
The rates for the decays $\B \to \jpsi \to e^+ e^-$ and $\B \to \psitwos \to e^+ e^-$
are also adjusted to~\cite{Yao:2007pdgupdate}.   

\begin{figure}[b]
\begin{center}
\includegraphics[height=1.8in]{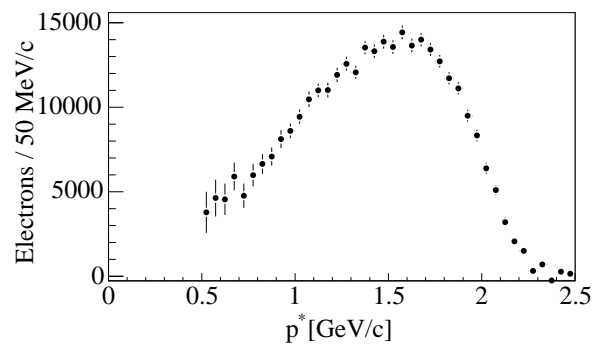} 
\caption {Electron-momentum spectrum from $\B \to X e \nu (\gamma)$ decays in the
$\Upsilon(4S)$ frame after correction for efficiencies and bremsstrahlung in the 
detector, with combined statistical and systematic errors.}
\label{fig_spectrum}
\end{center}
\end{figure}

After the like- and unlike-sign samples have been corrected for electron identification
efficiency, these irreducible background spectra are subtracted. 
To account for \BzBzb\ mixing, we determine the number of primary electrons in 
the $i$-th $p^*$-bin from the like-sign and unlike-sign pairs as
\begin{widetext}
\begin{equation}
N^i_{b \to c,u} = \frac{N^i_{\uls}}{\epsilon_{\alpha^*}^i} \frac{(1-f_0 \chi_0)-(1-\rho)(1-f_0)}{(1-2f_0 \chi_0)-(1-\rho)(1-f_0)(1-f_0 \chi_0)}
+ N^i_{\ls}\frac{f_0 \chi_0}{(1-2f_0 \chi_0)-(1-\rho)(1-f_0)(1-f_0 \chi_0)}
\label{eq:prompt}
\end{equation}
\end{widetext}
where $\chi_0 = 0.1878 \pm 0.0024$~\cite{Yao:2007pdgupdate} 
is the \BzBzb\ mixing parameter,
$f_0=\BR(\FourS \to \BzBzb)$ = $0.491\pm 0.007$~\cite{Yao:2007pdgupdate},
and
$\rho = \BR(B^+ \to \Dzb \to e^-) /  \BR(B^0 \to D^- \to e^-) = (0.744 \pm 0.06)$~\cite{Yao:2007pdgupdate}.
The parameter $\epsilon_{\alpha^*}^i$ is the efficiency of the additional
requirement for the  unlike-sign sample as defined in Eq.~(\ref{oacut}).
The spectrum obtained from Eq.~(\ref{eq:prompt}) is corrected for the 
effects of bremsstrahlung in the detector material using MC simulation.
Figure \ref{fig_spectrum} shows the resulting spectrum of primary 
electrons.

We determine the partial branching fraction as
$(\sum_{i}N^i_{b\to c,u})/(N_{\mathrm{tag}}\; \epsilon_{evt}\;\epsilon_{cuts})$,
where $i$ runs over all bins with $E_e > \emin$.
For the background-corrected number $N_{\mathrm{tag}}$ of tag electrons 
we find $N_{\mathrm{tag}} = (3617 \pm 4 \pm 22) \times 10^3$,
where the uncertainties are statistical and systematic, respectively.
The parameter $\epsilon_{evt}=(98.9 \pm 0.5)\%$ refers to the 
relative efficiency for selecting two-electron events compared 
to events with a single $e_{\mathrm{tag}}$, and
$\epsilon_{\mathrm{cuts}}=(82.8\pm 0.3)\%$ is the acceptance for the signal electron for 
$\emin=0.6\gev$. The result is
\begin{equation*}
\begin{aligned}
\BR(\B\to X e \nu (\gamma), & \;E_e>0.6\gev) & \\
       & = (10.30 \pm 0.06 \pm 0.21)\%,
\end{aligned}
\end{equation*}
where the errors correspond to the statistical and systematic uncertainties, respectively.

%

In the $\B$-meson rest frame, we define $R_i(\emin,\mu)$ 
as $\int_{\emin}^{\infty} (\Ee-\mu)^i (d\Gamma/dE_e)\,dE_e$,
and measure the first moment $M_1(\emin) = R_1(\emin,0) / R_0(\emin,0)$,
the central moments $M_n(\emin)=R_n(\emin,M_1(\emin))/R_0(\emin,0)$ for $n$
= 2, 3 and the partial branching fraction ${\cal B}(\emin) = \tau_B \; R_0(\emin,0)$, 
where $\tau_B$ is the average lifetime of charged and neutral $\B$ mesons. 
The calculation of the moments
is done as in ~\cite{Aubert:2004BABARLeptonMoments} and includes corrections
for charmless semileptonic decays, the movement of the $\B$ mesons in the 
c.m.~frame, biases due to the event selection criteria, and
binning effects. The spectra and moments presented are those of  
$\B \to X_c e \nu (\gamma)$ decays with any number of photons. Since current
theoretical predictions on the lepton-energy moments do not incorporate photon emission, 
we also present a second set of of moments with corrections for the impact 
of QED radiation using the $\photos$ code \cite{Richter-Was:1992PHOTOS}.

\begin{figure*}
\begin{center}
\includegraphics[height=2.3in]{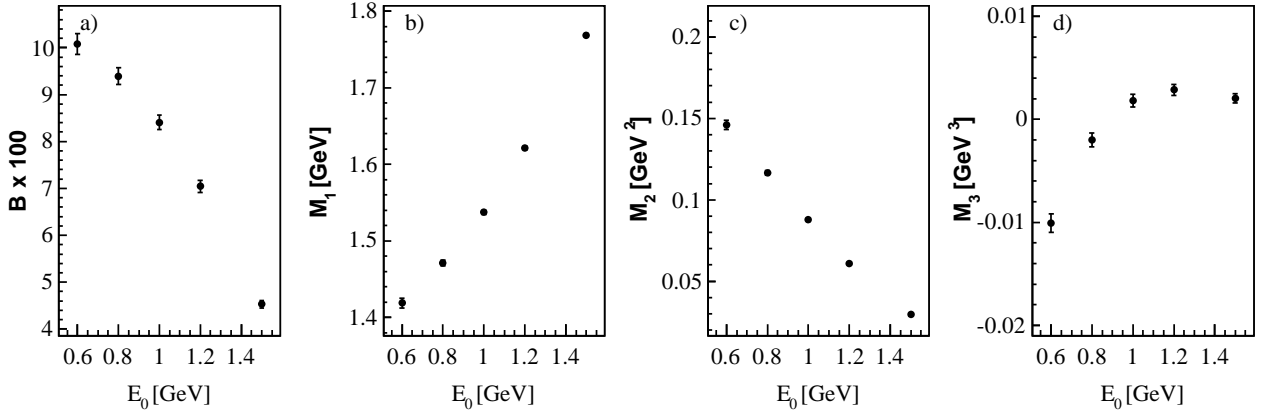}
\caption{Measured branching fraction (a) and moments 
$M_1$ (b), $M_2$ (c), and $M_3$ (d)
of the inclusive electron-energy 
spectrum of  $\B \to X_c e \nu (\gamma)$ decays as a function of the 
cutoff energy $E_0$ in the $\B$-meson rest frame.}
\label{fig:mom}
\end{center}
\end{figure*}

Figure \ref{fig:mom} shows the moments of $\B \to X_c e \nu (\gamma)$ decays
as a function of $\emin$, and Table~\ref{tab:sys} lists the main systematic 
errors for $\emin = 0.6$ and $1.5 \gev$. The complete listing 
of all moments and the full correlation matrix, with and without $\photos$ corrections
can be found in~\cite{ThisPRD:onlineversion}.

\section{Determination of $\Vcb$ and the quark masses $\mb$ and $\mc$}
\label{sec:fit}

At the parton level, the weak decay rate for $\b \rightarrow \c \ell \nu$  can be calculated
accurately; it is proportional to $\Vcb^2$ and depends on the quark masses
$m_b$ and $m_c$.  To relate measurements of the semileptonic $\B$-meson decay rate
to $\Vcb$, the parton-level calculations have to be corrected for effects of
strong interactions. Heavy Quark Expansions (HQEs)
\cite{Voloshin:1985HQE,Chay:1990HQE,Bigi:1991HQE}
have become a successful tool for calculating perturbative and nonperturbative QCD corrections
\cite{Bigi:1993PRLB293HQE,Bigi:1993PRL71HQE,Blok:1994HQE,Manohar:1994HQE,Gremm:1997HQE}
and for estimating their uncertainties.

In the kinetic-mass scheme 
\cite{Benson:2003GammaKineticScheme,Gambino:2004MomentsKineticScheme,
  Benson:2004BToSGammaKineticScheme,Aquila:2005PertCorrKineticScheme,
  Uraltsev:2004PertCorrKineticScheme,Bigi:2005KineticSchemeOpenCharm},
these expansions in $1/m_b$ and the strong coupling constant $\alpha_s(m_b)$ 
to order ${\cal O}(1/m_b^3)$ contain six parameters: the running kinetic masses of the $b$ and
$c$ quarks, $\mb(\mu)$ and $\mc(\mu)$, and four nonperturbative parameters.
The parameter $\mu$ denotes the Wilson factorization scale that separates
effects from long- and short-distance dynamics. The calculations are performed
for $\mu = 1 \gev$ \cite{1997:BigiKineticScheme}. It has been shown that the expressions 
for the moments have only a small scale dependence \cite{Buchmuller:2005globalhqefit}.
We determine these six parameters and $\Vcb$ from fits to moments of the hadronic-mass,
combined mass-and-energy, and electron-energy distributions in semileptonic $\B$ 
decays $\semilepXc$ and moments of the photon-energy spectrum in decays $\BtoXsGamma$
\cite{Aubert:2005BABARXsGammaExclusive,Aubert:2006XsGammaInclusive, Aubert:2007XsGammaBReco}.

In the kinetic-mass scheme the HQE to ${\cal O}(1/m_b^3)$ for the rate $\Gammasl$ of
semileptonic decays $\semilepXc$ can be expressed as \cite{Benson:2003GammaKineticScheme}

\begin{eqnarray}
   \Gammasl & = & \frac{G_F^2 \mb^5}{192\pi^3} \Vcb^2 (1+A_{\mathit{ew}}) A_{\mathit{pert}}(r,\mu)  \nonumber\\
      & \times & \Bigg [ z_0(r) \Bigg ( 1 - \frac{\mupi - \muG +
            \frac{\rhoD + \rhoLS}{c^2 \mb}}{2 c^4 \mb^2} \Bigg )  \\
      & - & 2(1-r)^4\frac{\muG + \frac{\rhoD + \rhoLS}{c^2 \mb}}{c^4 \mb^2}
        + d(r)\frac{\rhoD}{c^6 \mb^3} \nonumber \\
      & + & \mathcal{O}(1/\mb^4)\Bigg]. \nonumber
   \label{eq:vcb_gammaslkinetic}
\end{eqnarray}

The leading nonperturbative effects arise at ${\cal O}(1/\mb^2)$ and are parameterized by
$\mupi(\mu)$ and $\muG(\mu)$, the expectation values of the kinetic and
chromomagnetic dimension-five operators. At ${\cal O}(1/\mb^3)$, two additional parameters enter,
$\rhoD(\mu)$ and $\rhoLS(\mu)$, the  expectation values of the Darwin and
spin-orbit dimension-six operators, respectively.
The ratio $r = m_c^2/m_b^2$ enters in the tree level phase space factor
$z_0(r) = 1 - 8r + 8r^3 - r^4 - 12r^2 \ln r$ and in the function
$d(r) = 8 \ln r + 34/3 - 32r/3 - 8r^2 + 32r^3/3 - 10r^4 /3$.
The factor $1 + A_{\mathit{ew}}$ accounts for electroweak corrections. It
is estimated to be $1 + A_{ew} \cong ( 1 + \alpha/\pi \ln M_Z/\mb )^2 = 1.014$,
where $\alpha$ is the electromagnetic coupling constant.
The quantity $A_{\mathit{pert}}$ accounts for perturbative contributions and
is estimated to be $A_{pert}(r,\mu) \approx 0.908$ \cite{Benson:2003GammaKineticScheme}.

\begin{table*} 
\caption{Results and breakdown of the systematic errors for the partial branching fraction
$\BR = \tau_B \int_{\emin}^{\infty} (d\Gamma/dE_e)\,dE_e$, and the moments $M_1$, $M_2$, and $M_3$
for $B\to X_c e\nu$ and $B\to X_c e\nu (\gamma)$ in the $\B$-meson rest 
frame for two values of \emin. Changes wrt.~the previously published results \cite{Aubert:2004BABARLeptonMoments}
are due to updated background branching fractions \cite{Yao:2007pdgupdate, BABAR:BtoDUppperVertex}
and indicated by ($\dagger$).}
\begin{center} 

\begin{tabular}{lcccccccc}
\hline 
\hline 
 &\multicolumn{2}{c}{${\cal B} [10^{-2}]$} &\multicolumn{2}{c}{$M_1 [\mev]$} &\multicolumn{2}{c}{$M_2 [10^{-3} \gev^2]$} &\multicolumn{2}{c}{$M_3 [10^{-3} \gev^3]$}  \\
$E_{0} [\gev]$ &0.6 &1.5 &0.6 &1.5 &0.6 &1.5 &0.6 &1.5  \\
\hline 
\multicolumn{9}{c}{Breakdown of systematic errors} \\
\hline
Conversion and Dalitz pairs &$0.028$ &$0.001$ &$1.5$ &$0.02$ &$0.6$ &$0.00$ &$0.04$ &$0.00$  \\
$e$ identification efficiency &$0.150$ &$0.044$ &$2.5$ &$0.30$ &$0.6$ &$0.07$ &$0.27$ &$0.08$  \\
$e$ from same $B$ &$0.019$ &$0.000$ &$1.3$ &$0.00$ &$0.6$ &$0.00$ &$0.05$ &$0.00$  \\
$\B \to D_s \to e$ $(\dagger)$ &$0.024$ &$0.000$ &$1.4$ &$0.01$ &$0.5$ &$0.00$ &$0.03$ &$0.00$  \\
$\B \to D \to e$ $(\dagger)$ &$0.035$ &$0.000$ &$2.2$ &$0.00$ &$1.0$ &$0.00$ &$0.03$ &$0.00$  \\
$\B \to \tau \to e$ $(\dagger)$ &$0.027$ &$0.001$ &$1.2$ &$0.04$ &$0.3$ &$0.00$ &$0.10$ &$0.00$  \\
$e$ from $J/\psi$ or $\psi(2S)$ $(\dagger)$ &$0.002$ &$0.001$ &$0.0$ &$0.01$ &$0.0$ &$0.01$ &$0.00$ &$0.00$  \\
Secondary tags &$0.052$ &$0.011$ &$1.6$ &$0.06$ &$0.6$ &$0.00$ &$0.05$ &$0.00$  \\
$\chi$ &$0.022$ &$0.011$ &$0.9$ &$0.01$ &$0.3$ &$0.00$ &$0.03$ &$0.00$  \\
Tracking efficiency &$0.083$ &$0.033$ &$1.0$ &$0.06$ &$0.3$ &$0.02$ &$0.06$ &$0.00$  \\
Bremsstrahlung correction &$0.011$ &$0.028$ &$1.9$ &$0.43$ &$0.0$ &$0.05$ &$0.18$ &$0.00$  \\
Event selection &$0.052$ &$0.024$ &$0.6$ &$0.14$ &$0.0$ &$0.03$ &$0.07$ &$0.01$  \\
$b \to u$ subtraction $(\dagger)$ &$0.031$ &$0.020$ &$0.8$ &$0.83$ &$0.4$ &$0.32$ &$0.14$ &$0.12$  \\
$B$ momentum correction &$0.000$ &$0.005$ &$0.0$ &$0.19$ &$0.1$ &$0.10$ &$0.04$ &$0.02$  \\
$N_{tag}$ normalization &0.068 &0.030 & & & & & &  \\
\hline 
\multicolumn{9}{c}{Results} \\
\hline 
Results for $B\to X_c e\nu (\gamma)$ &$10.08$ &$4.53$ &$1418.8$ &$1768.7$ &$146.1$ &$29.6$ &$-10.08$ &$2.04$  \\
$\pm$(stat.) &$0.06$ &$0.03$ &$3.8$ &$1.9$ &$2.0$ &$0.8$ &$0.81$ &$0.44$  \\
$\pm$(syst.) &$0.21$ &$0.08$ &$5.4$ &$1.0$ &$1.9$ &$0.4$ &$0.40$ &$0.15$  \\
\hline 
Results for $B\to X_c e\nu$ &$10.20$ &$4.78$ &$1437.6$ &$1773.7$ &$145.4$ &$30.1$ &$-12.04$ &$2.04$  \\
$\pm$(stat.) &\ $0.06$\  &\ $0.03$\  &\ $4.0$\  &\ $1.9$\  &\ $2.3$\  &\ $0.9$\  &\ $0.91$\  &\ $0.47$\   \\
$\pm$(syst.) &\ $0.22$\  &\ $0.08$\  &\ $5.7$\  &\ $1.1$\  &\ $2.1$\  &\ $0.4$\  &\ $0.40$\  &\ $0.15$\   \\
\hline 
\hline 
\end{tabular} 

\label{tab:sys}
\end{center} 
\end{table*}

The performed fit uses a linearized expression for the dependence
of $\Vcb$ on the values of heavy-quark parameters, expanded around
${\it a~priori}$ estimates of these parameters \cite{Benson:2003GammaKineticScheme}:
\begin{eqnarray}
   \frac{\Vcb}{0.0417}  & & = \sqrt{\frac{\brf(\semilepXc)}{0.1032} \frac{1.55}{\tau_{\B}} } \\
                        & &\times [1 + 0.30 (\alpha_s(\mb) - 0.22) ] \nonumber \\
                        & &\times [ 1 - 0.66 ( \mb - 4.60) + 0.39 ( \mc - 1.15 ) \nonumber\\
                        & &+ 0.013 ( \mupi - 0.40) + 0.09 ( \rhoD - 0.20) \nonumber\\
                        & &+ 0.05 ( \muG - 0.35 ) - 0.01 ( \rhoLS + 0.15 ) ]. \nonumber
   \label{eq:vcb_gammaslkineticlinear}
\end{eqnarray}
Here $\mb$ and $\mc$ are in $\gevcc$ and all other parameters of the expansion are in
$\gev^{k}$; $\tau_B$ refers to the average lifetime of $\B$ mesons produced at the $\FourS$, measured in picoseconds.
HQEs in terms of the same heavy-quark parameters are available for hadronic-mass,
combined mass-and-energy, electron-energy, and photon-energy moments. 
Predictions for those moments are obtained
from an analytical calculation \cite{Uraltsev:private}. We use these calculations to determine 
$\Vcb$, the total semileptonic branching fraction $\brf(\semilepXc)$, 
the quark masses $\mb$ and $\mc$,
as well as the heavy-quark parameters $\mupi$, $\muG$, $\rhoD$, and $\rhoLS$,
from a simultaneous $\chisq$ fit to the measured moments and partial branching fractions,
all as functions of the minimum lepton momentum $\plmin$ and minimum photon energy 
$\Egammacut$.

\subsection{Extraction Formalism}

The fit method designed to extract the HQE parameters from the measured moments 
has been reported previously \cite{Buchmuller:2005globalhqefit, Aubert:2004BABARHQEFit}.
It is based on a $\chisq$ minimization,
\begin{equation}
    \chisq  = 
   \left( \vec{M}_{\mathrm{exp}} - \vec{M}_{\mathrm{HQE}} \right)^{T}
               \covtot^{-1}
             \left( \vec{M}_{\mathrm{exp}} - \vec{M}_{\mathrm{HQE}} \right).
    \label{eq:vcb_chi2}
\end{equation}
The vectors $\vec{M}_{\mathrm{exp}}$ and $\vec{M}_{\mathrm{HQE}}$ contain the measured
moments and the corresponding moments calculated by theory,
respectively. Furthermore, the expression in Eq.~(\ref{eq:vcb_chi2})
contains the total covariance matrix $\covtot = \covexp +  \covhqe$ 
defined as the sum of the experimental $\covexp$ and theoretical $\covhqe$
covariance matrices (see Section \ref{sec:vcb_theoreticalerrors}).

The total semileptonic branching fraction $\brf(\semilepXc)$ is extracted
in the fit by extrapolating the measured partial branching fractions
$\brf_{\plcut}(\semilepXc)$ with $\plep \geq \plmin$ to the full lepton
energy spectrum. Using HQE predictions of the relative decay fraction
\begin{equation}
    R_{\plcut} = \frac{\int_{\plcut} \frac{\text{d}\Gammasl}{\text{d}\plep} \text{d} \plep}
                      {\int_{0} \frac{\text{d}\Gammasl}{\text{d}\plep} \text{d} \plep},
\end{equation}
the total branching fraction can be introduced as a free parameter in the fit.
It is given by
\begin{equation}
    \brf(\semilepXc) = \frac{\brf_{\plcut}(\semilepXc)}{R_{\plcut}}.
\end{equation}
Using Eqs.~(\ref{eq:vcb_gammaslkinetic}) and (\ref{eq:vcb_gammaslkineticlinear})
together with the measured average $\B$-meson lifetime $\tau_{\B}$
and the total branching fraction, allows the calculation of
$\Vcb$:
\begin{equation}
    \Vcb^{2} \propto \Gammasl = \frac{\brf(\semilepXc)}{\tau_{\B}}.
\end{equation}
Thereby, $\Vcb$ is introduced as an additional free parameter to the fit.
To propagate the uncertainty on $\tau_{\B}$ properly into the extracted
result for $\Vcb$, $\tau_{\B}$ is added as an additional measurement to the
vectors of measured and predicted quantities, $\vec{M}_{\mathrm{exp}}$
and $\vec{M}_{\mathrm{HQE}}$.

The nonperturbative parameters $\muG$ and $\rhoLS$ have been estimated from the
$\B$-$\B^*$ mass splitting and heavy-quark sum rules to be
$\muG = (0.35 \pm 0.07) \gev^{2}$ and $\rhoLS = (-0.15 \pm 0.10) \gev^{3}$
\cite{Buchmuller:2005globalhqefit}, respectively. Both parameters are restricted
in the fit by imposing Gaussian error constraints.

\subsection{Experimental Input}

The combined fit is performed on a subset of available moment measurements
with correlations below $95\%$ to ensure the invertibility of the covariance matrix.
Since the omitted measurements are characterized by high correlations to other
measurements considered in the fit, they do not contribute significant additional
information, and the overall sensitivity of the results is not affected.
Choosing a different subset of moments gives consistent results.
We perform two fits to the following set of measured moments, thereby including either
the hadronic-mass moments or the moments of the combined mass-and-energy spectrum:
\begin{itemize}

   \item Hadronic-mass moments are used as presented in this paper. 
         We select the following subset for the fit: 
         $\mxmom{2}$ for  $\plgeq{0.9,1.1,1.3,1.5}$, 
         $\mxmom{4}$ for $\plgeq{0.8,1.0,1.2, 1.4}$, 
         and $\mxmom{6}$ for $\plgeq{0.9,1.1,1.3,1.5}$.

   \item Moments of the combined mass-and-energy spectrum as presented in
         this paper. The following subset of moments is included in the fit:
         $\moment{\nx}$ for $\plgeq{0.9,1.1,1.3,1.5}$, 
         $\moment{\nxfour}$ for $\plgeq{0.8,1.0,1.2, 1.4}$, 
         and $\moment{\nxsix}$ for $\plgeq{0.9,1.1,1.3,1.5}$.
\end{itemize}
Both fits include the updated lepton-energy moments as presented in this paper 
with radiative corrections as well as photon-energy moments measured in 
$\BtoXsGamma$ decays as presented in 
\cite{Aubert:2005BABARXsGammaExclusive, Aubert:2006XsGammaInclusive, 
Aubert:2007XsGammaBReco}.
We use the partial branching fraction $\brf_{\plcut}$ measured for
$\plgeq{0.6,1.0,1.5}$ and the moments $\elmom{}$ for $\plgeq{0.6,0.8,1.0,1.2,1.5}$. 
The lepton-energy moments $\elmom{2}$ are used 
for the minimum lepton momentum $\plgeq{0.6,1.0,1.5}$ and $\elmom{3}$ for $\plgeq{0.8,1.2}$.
We include the photon-energy moments $\egammamom{}$ for the minimum photon energies
$\Egammageq{1.9}$ and $\Egammageq{2.0}$, and $\egammamom{2}$ for $\Egammageq{1.9}$.
In addition, we use $\tau_{\B} = f_0 \tau_0 + (1 - f_0) \tau_{\pm} = (1.585 \pm 0.007) \ps$,
taking into account the lifetimes \cite{Yao:2007pdgupdate} of neutral and charged $\B$ mesons,
$\tau_0$ and $\tau_{\pm}$, and their relative production rates,
$f_0 = 0.491 \pm 0.007$ \cite{Yao:2007pdgupdate}.

\subsection{Theoretical Uncertainties}
\label{sec:vcb_theoreticalerrors}

As discussed in \cite{Buchmuller:2005globalhqefit} and specified in
\cite{Gambino:2004MomentsKineticScheme}, the following theoretical
uncertainties are taken into account:

The uncertainty related to the uncalculated perturbative corrections
to the Wilson coefficients of nonperturbative operators
are estimated by varying the corresponding parameters
$\mupi$ and $\muG$ by $20\%$ and $\rhoD$ and $\rhoLS$
by $30\%$ around their expected values.
Uncertainties for the perturbative corrections are
estimated by varying $\alpha_{s}$ up and down by $0.1$ for the hadronic-mass moments
and by $0.04$ for the lepton-energy moments around its nominal value of $\alpha_{s} = 0.22$.
Uncertainties in the perturbative corrections of the quark masses $m_{\b}$
and $m_{c}$ are addressed by varying both by $20\mevcc$ up and down around
their expected values.

For the extracted value of $\Vcb$ an additional error
of $1.4\%$ is added for the uncertainty in the expansion of the
semileptonic rate $\Gammasl$
\cite{Benson:2003GammaKineticScheme, Bigi:2005KineticSchemeOpenCharm}.
It accounts for remaining uncertainties in the perturbative corrections
to the leading operator, uncalculated perturbative corrections
to the chromomagnetic and Darwin operator, higher order power corrections, and
possible nonperturbative effects in the operators with charm fields.
This uncertainty is not included in the theoretical covariance matrix
\covhqe\ but is listed separately as a theoretical uncertainty on $\Vcb$.

For the predicted photon-energy moments $\egammamom{n}$, additional
uncertainties are taken into account. As outlined in \cite{Benson:2004BToSGammaKineticScheme},
uncertainties of $30\%$ of the applied bias correction
to the photon-energy moments and half the difference in the moments
derived from two different distribution-function \textit{ans\"atze} have to be
considered. Both contributions are added linearly \cite{Buchmuller:2005globalhqefit}.

The theoretical covariance matrix $\covhqe$ is constructed by assuming
fully correlated theoretical uncertainties for a given moment with different
lepton-momentum or photon-energy cutoffs and assuming uncorrelated theoretical 
uncertainties for moments of different orders and types. The additional 
uncertainties considered for the photon-energy moments are assumed to be 
uncorrelated for different moments and photon-energy cutoffs.

\subsection{Results}

In the following, the results of the two fits, one including
the measurement of hadronic-mass moments and the other
including the measured moments of the combined mass-and-energy
spectrum instead, are discussed. 

We use a parameterized MC simulation to separate fit parameter uncertainties
into experimental and theoretical contributions. 
The simulation uses a set of expected moments randomly varied with
either $\covtot$ or $\covexp$. Fits to these moments allow
for the determination of the expected total and experimental uncertainties,
respectively. The final experimental and theoretical uncertainties are
calculated from the final total uncertainties by means of their simulated
relative expected fractions.

\subsubsection{Combined Fit Including Hadronic-Mass Moments}

\begin{figure*}
  \begin{center}
  \includegraphics[width=0.85\textwidth]{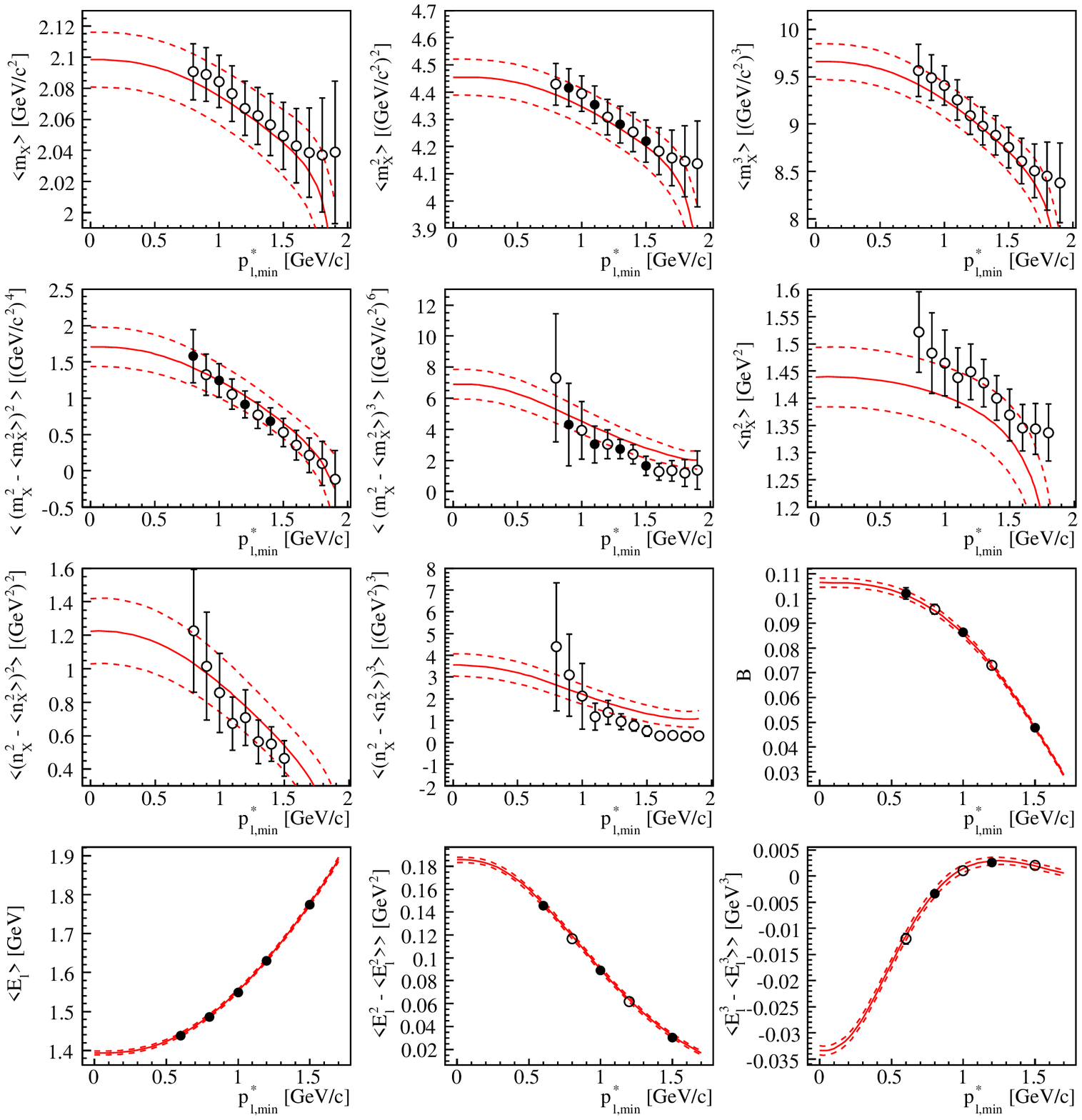}
  \includegraphics[width=0.85\textwidth]{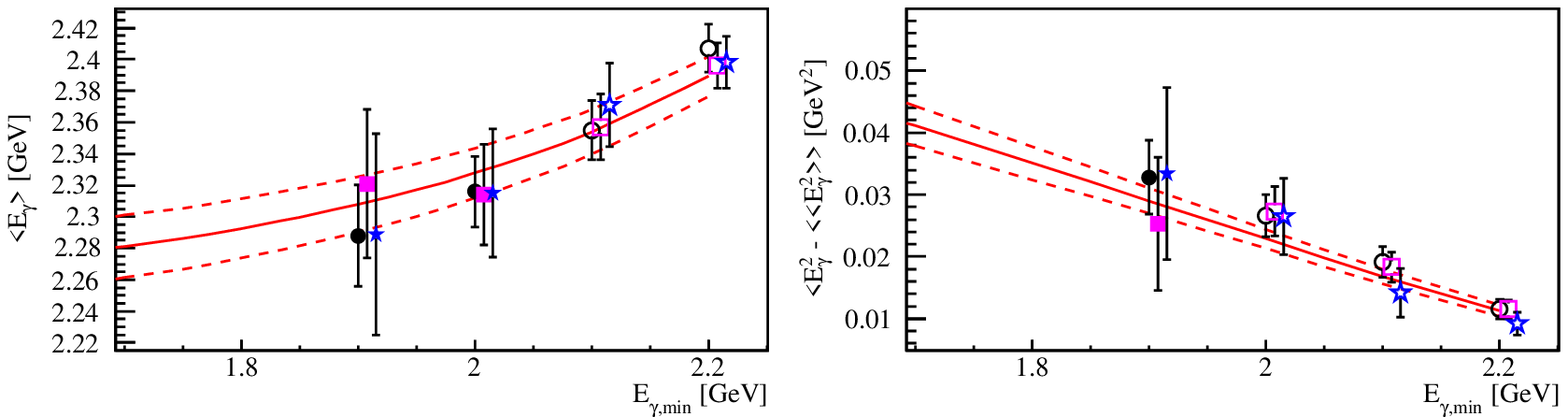}
  \end{center}
  \caption{The measured hadronic-mass moments $\mxmom{k}$, combined 
           mass-and-energy moments $\moment{\nxn}$,
           electron-energy moments $\elmom{k}$, partial branching fractions B, and
           photon-energy moments $\egammamom{n}$,
           as a function of the minimum lepton momentum $\plmin$ 
           and minimum photon energy $E_{\gamma,min}$
           compared with the result of the 
           simultaneous fit (solid line) to moments of the hadronic 
           mass spectrum, electron-energy moments, and photon-energy moments.
           The solid data points mark the measurements included in the fit. 
           Moments of semileptonic decays $\semilepXc$ are marked by (\textcolor{black}{$\bullet$}). 
           Photon-energy moments of Ref.~\cite{Aubert:2005BABARXsGammaExclusive} are marked 
           by (\textcolor{magenta}{$\blacksquare$}), of Ref.~\cite{Aubert:2006XsGammaInclusive} 
           by (\textcolor{black}{$\bullet$}), and of Ref.~\cite{Aubert:2007XsGammaBReco} 
           by (\textcolor{blue}{$\bigstar$}).
           Open data points are not used in the fit. 
           The vertical bars indicate the experimental errors.
           The dashed lines correspond to the total fit uncertainty as
           obtained by converting the fit errors of each individual HQE parameter
           into an error for the individual moment.
          }
   \label{fig:vcb_FitMassMoments}
\end{figure*}

A comparison of the fit including hadronic-mass moments
with the measured moments is shown in Fig.~\ref{fig:vcb_FitMassMoments}.
The moments \mxmom{} and \mxmom{3} as well as the combined mass-and-energy moments
are not included in the fit and thus provide an unbiased comparison
with the fitted HQE prediction. We find an overall good agreement,
also indicated by $\chisq = \resultfitchisqmx$ for $28$ degrees of freedom.
Results for the SM and HQE parameters extracted from the
fit are summarized in Table \ref{tab:vcb_fitResults_Mx}. We find
$\Vcb = (42.05 \pm 0.45 \pm 0.70) \times 10^{-3}$, 
$\BR(\Bbar \to X_c \en \bar{\nu}) = (10.64 \pm 0.17 \pm 0.06)\%$,
$\mb = (4.549 \pm 0.031 \pm 0.038) \gevcc$, 
and $\mc = (1.077 \pm 0.041 \pm 0.062) \gevcc$, where the errors correspond to 
experimental and theoretical uncertainties, respectively. 
The fitted quark masses have a large correlation of $95\%$ 
resulting in a more precise determination of the
quark mass difference, $\mb - \mc = (3.472 \pm 0.032) \gevcc$, where the error
is the total uncertainty.
We translate the quark masses which were extracted in the kinetic scheme
into the $\overline{\mathrm{MS}}$ scheme using calculations
up to $\mathcal{O}(\alpha_{s}^{2})$ accuracy \cite{Benson:2003GammaKineticScheme}.
The translation yields $\mbbar(\mbbar) = (4.186 \pm 0.044 \pm 0.015) \gevcc$ and
$\mcbar(\mcbar) = (1.196 \pm 0.059 \pm 0.050) \gevcc$, where the first uncertainty
is a translation of the uncertainty obtained in the kinetic scheme
and the second corresponds to an estimate for the uncertainty of the
transformation itself.

\subsubsection{Combined Fit Including Combined Mass-and-Energy Moments}

\begin{figure*}
  \begin{center}
  \includegraphics[width=0.85\textwidth]{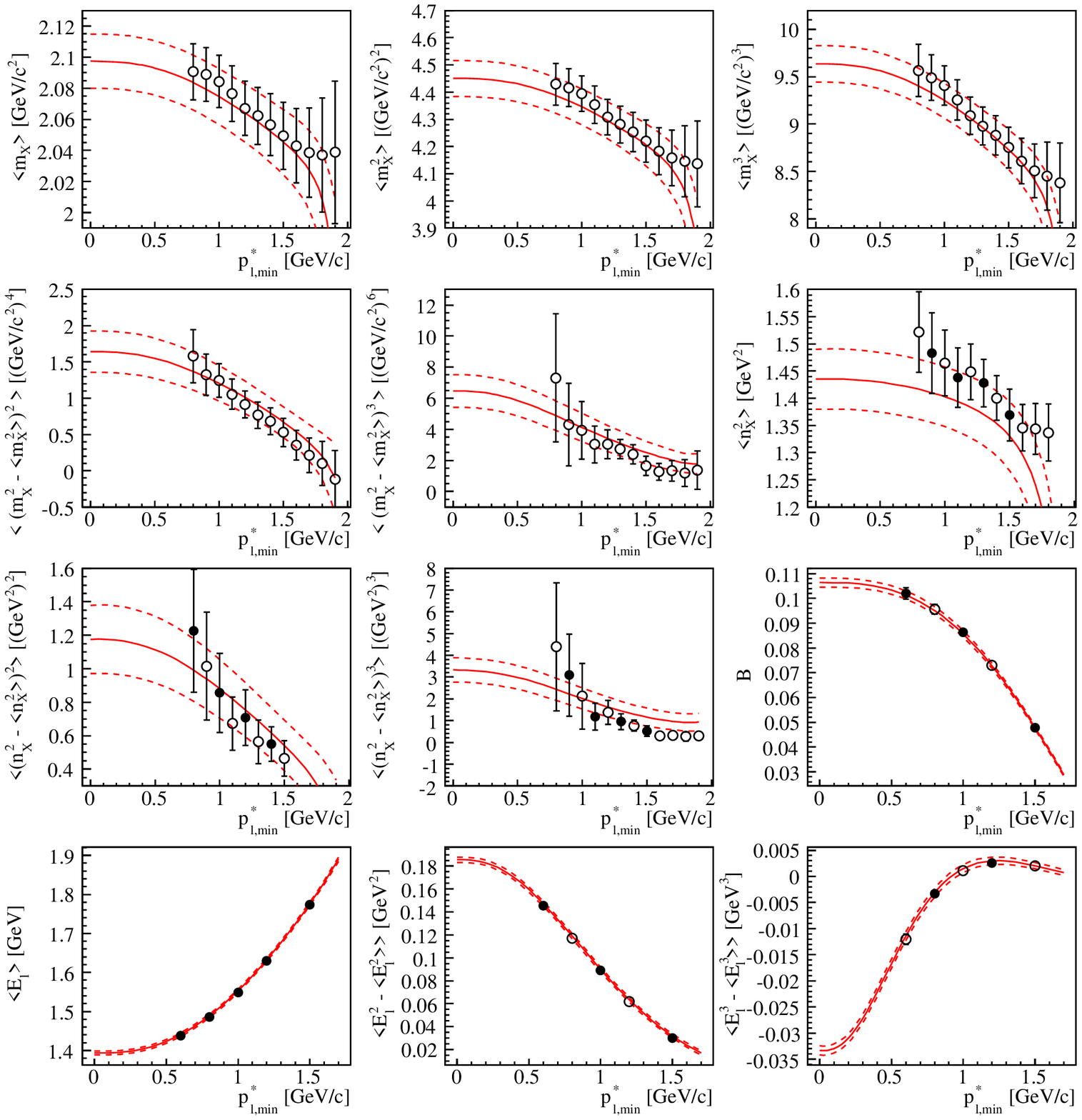}
  \includegraphics[width=0.85\textwidth]{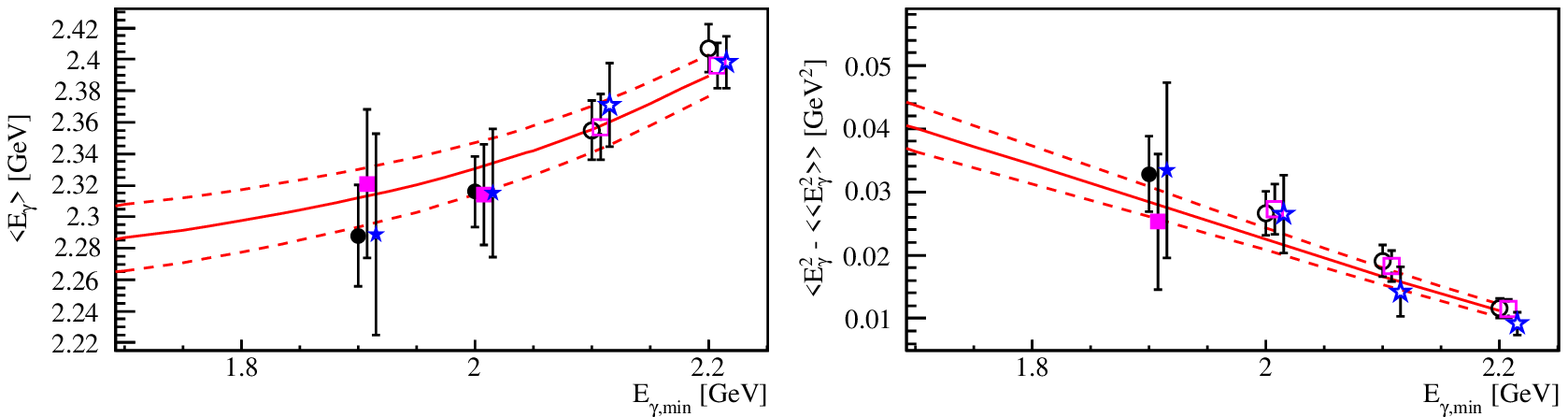}
  \end{center}
  \caption{The measured hadronic-mass moments $\mxmom{k}$, 
           combined mass-and-energy moments $\moment{\nxn}$,
           electron-energy moments $\elmom{k}$, 
           partial branching fractions B, and
           photon-energy moments $\egammamom{n}$,
           as a function of the minimum lepton momentum $\plmin$ 
           and minimum photon energy $E_{\gamma,min}$
           compared with the result of the simultaneous fit (solid line) to 
           moments of the combined mass-and-energy spectrum,
           electron-energy moments, and photon-energy moments.
           The solid data points mark the measurements included in the fit. 
           Moments of semileptonic decays $\semilepXc$ are marked 
           by (\textcolor{black}{$\bullet$}). 
           Photon-energy moments of Ref.~\cite{Aubert:2005BABARXsGammaExclusive} are marked 
           by (\textcolor{magenta}{$\blacksquare$}), of Ref.~\cite{Aubert:2006XsGammaInclusive} 
           by (\textcolor{black}{$\bullet$}), and of Ref.~\cite{Aubert:2007XsGammaBReco} 
           by (\textcolor{blue}{$\bigstar$}).
           Open data points are not used in the fit. 
           The vertical bars indicate the experimental errors.
           The dashed lines correspond to the total fit uncertainty as
           obtained by converting the fit errors of each individual HQE parameter
           into an error for the individual moment.
          }
   \label{fig:vcb_FitMixedMoments}
\end{figure*}

Figure \ref{fig:vcb_FitMixedMoments} shows a comparison 
of the measured moments and the fit including 
the measured combined mass-and-energy moments. We find an overall good agreement with
$\chisq = \resultfitchisqnx$ for $28$ degrees of freedom.
The fit yields predictions of the hadronic-mass moments that are
in good agreement with the measurement.
Numerical results of the fit are summarized in Table \ref{tab:vcb_fitResults_Nx}.
We find $\Vcb = (41.91 \pm 0.48 \pm 0.70) \times 10^{-3}$, 
$\BR(\Bbar \to X_c \en \bar{\nu}) = (10.64 \pm 0.17 \pm 0.06)\%$,
$\mb = (4.566 \pm 0.034 \pm 0.041) \gevcc$,
and $\mc = (1.101 \pm 0.045 \pm 0.064) \gevcc$, where the errors correspond to 
experimental and theoretical uncertainties, respectively. 
The two masses are correlated with $95\%$. 
Their difference is $\mb - \mc = (3.465 \pm 0.032) \gevcc$, where
the stated uncertainty corresponds to the total uncertainty.
The extracted quark masses translate into the $\overline{\mathrm{MS}}$ scheme
using \cite{Benson:2003GammaKineticScheme} as 
$\mbbar(\mbbar) = (4.201 \pm 0.047 \pm 0.015) \gevcc$ and
$\mcbar(\mcbar) = (1.215 \pm 0.062 \pm 0.050) \gevcc$, where the first uncertainty
is a translation of the uncertainty obtained in the kinetic scheme
and the second corresponds to an estimate for the uncertainty of the
transformation itself.

\begin{table*}
\caption{Results of the simultaneous fit to moments of the hadronic-mass spectrum,
         electron-energy moments, and photon-energy moments.
         For $\Vcb$ we account for an additional theoretical uncertainty
         of $1.4\%$ from the uncertainty in the expansion of $\Gammasl$.
         Correlation coefficients for all parameters are summarized below the central values.
       }

\begin{tabular}{lrrrrrrrr}
\hline \hline
 &$\Vcb \, \times 10^{3}$ &$\mb \, [\gevcc]$ &$\mc \, [\gevcc]$ &$\brf \, [\%]$ &$\mupi \, [\gev^{2}]$ &$\muG \, [\gev^{2}]$ &$\rhoD \, [\gev^{3}]$ &$\rhoLS \, [\gev^{3}]$  \\
\hline 
Results &42.05 &4.549 &1.077 &10.642 &0.476 &0.300 &0.203 &-0.144  \\
$\Delta_{exp}$ &0.45 &0.031 &0.041 &0.165 &0.021 &0.044 &0.017 &0.075  \\
$\Delta_{theo}$ &0.37 &0.038 &0.062 &0.063 &0.059 &0.038 &0.027 &0.056  \\
$\Delta_{\Gammasl}$ &0.59 & & & & & & &  \\
$\Delta_{tot}$ &0.83 &0.049 &0.074 &0.176 &0.063 &0.058 &0.032 &0.094  \\
\hline 
$\Vcb$ &1.00 &-0.33 &-0.11 &0.76 &0.32 &-0.42 &0.40 &0.12  \\
$\mb$ & &1.00 &0.95 &0.08 &-0.52 &0.14 &-0.22 &-0.24  \\
$\mc$ & & &1.00 &0.15 &-0.56 &-0.12 &-0.21 &-0.15  \\
$\brf$ & & & &1.00 &0.16 &-0.09 &0.16 &-0.06  \\
$\mupi$ & & & & &1.00 &0.04 &0.62 &0.08  \\
$\muG$ & & & & & &1.00 &-0.08 &-0.04  \\
$\rhoD$ & & & & & & &1.00 &-0.14  \\
$\rhoLS$ & & & & & & & &1.00  \\
\hline 
\hline 
\end{tabular} 

\label{tab:vcb_fitResults_Mx}
\end{table*}

\begin{table*}
\caption{Results of the simultaneous fit to moments of the combined 
         mass-and-energy spectrum, electron-energy moments, and photon-energy moments.
         For $\Vcb$ we account for an additional theoretical uncertainty
         of $1.4\%$ from the uncertainty in the expansion of $\Gammasl$.
         Correlation coefficients for all parameters are summarized below the central values.
       }

\begin{tabular}{lrrrrrrrr}
\hline \hline
 &$\Vcb \, \times 10^{3}$ &$\mb \, [\gevcc]$ &$\mc \, [\gevcc]$ &$\brf \, [\%]$ &$\mupi \, [\gev^{2}]$ &$\muG \, [\gev^{2}]$ &$\rhoD \, [\gev^{3}]$ &$\rhoLS \, [\gev^{3}]$  \\
\hline 
Results &41.91 &4.566 &1.101 &10.637 &0.452 &0.304 &0.190 &-0.156  \\
$\Delta_{exp}$ &0.48 &0.034 &0.045 &0.166 &0.023 &0.047 &0.013 &0.079  \\
$\Delta_{theo}$ &0.38 &0.041 &0.064 &0.061 &0.065 &0.039 &0.031 &0.052  \\
$\Delta_{\Gammasl}$ &0.59 & & & & & & &  \\
$\Delta_{tot}$ &0.85 &0.053 &0.078 &0.176 &0.069 &0.061 &0.034 &0.095  \\
\hline 
$\Vcb$ &1.00 &-0.43 &-0.24 &0.74 &0.41 &-0.43 &0.43 &0.15  \\
$\mb$ & &1.00 &0.95 &0.04 &-0.58 &0.20 &-0.30 &-0.27  \\
$\mc$ & & &1.00 &0.11 &-0.62 &-0.05 &-0.30 &-0.19  \\
$\brf$ & & & &1.00 &0.17 &-0.09 &0.16 &-0.05  \\
$\mupi$ & & & & &1.00 &0.01 &0.68 &0.14  \\
$\muG$ & & & & & &1.00 &-0.05 &-0.05  \\
$\rhoD$ & & & & & & &1.00 &-0.08  \\
$\rhoLS$ & & & & & & & &1.00  \\
\hline 
\hline 
\end{tabular} 

\label{tab:vcb_fitResults_Nx}
\end{table*}

\subsubsection{Comparison of Results}

Comparing the result of the fit that includes moments of the $\nx$ distribution
with that including hadronic-mass moments instead, we find good agreement
of all fit parameters and their uncertainties. The differences between 
the fit values are $0.2$ $\sigma$ for $\Vcb$, $0.3$ $\sigma$ for $\mb$, 
and $0.3$ $\sigma$ for $\mc$. The uncertainties of all fit parameters  
in both fits agree within $8\%$.

\begin{figure*}
  \begin{center}
  \includegraphics[width=0.9\textwidth]{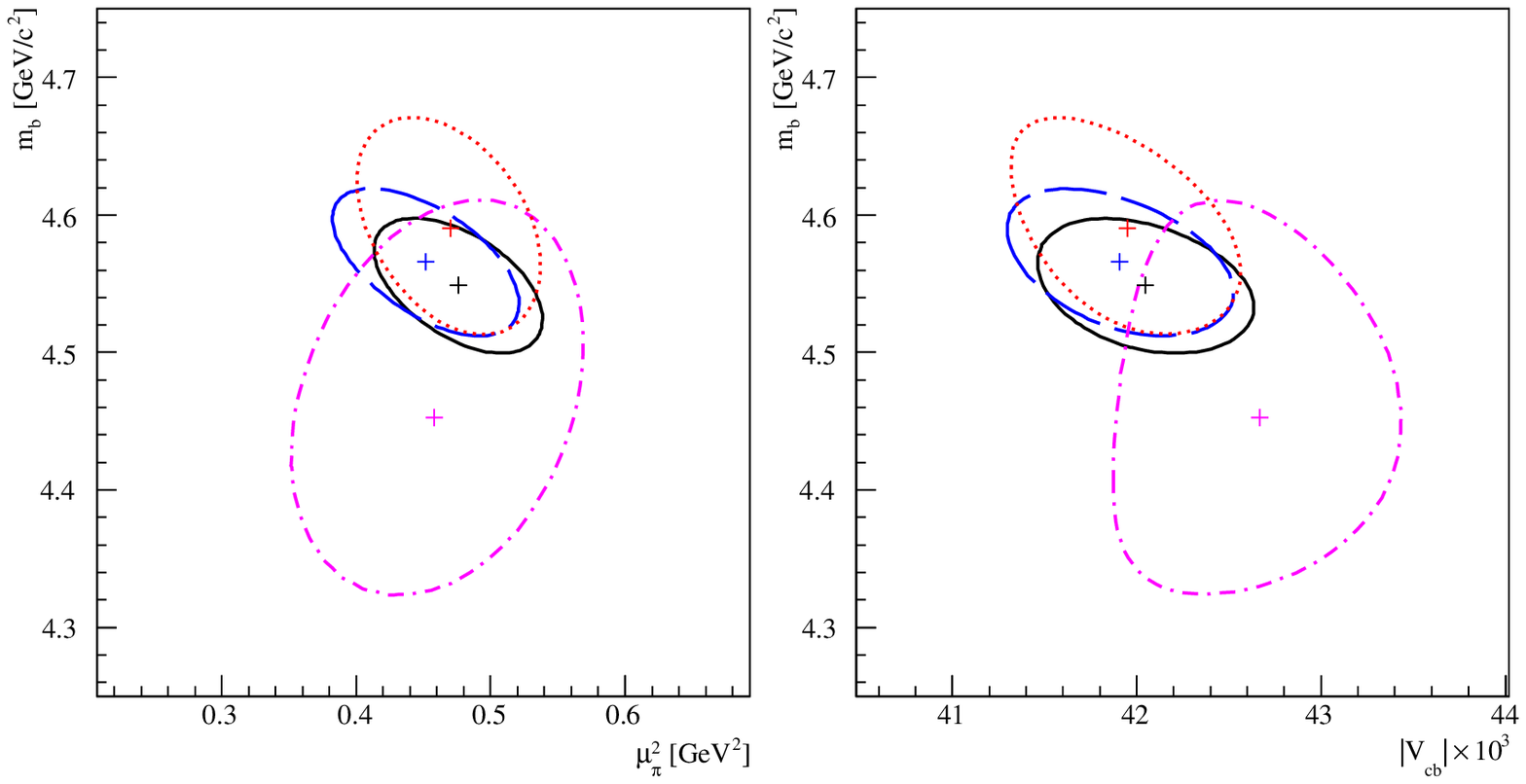}
  \end{center}
      \caption{$\Delta\chisq = 1$ contours for different fits 
               in the $(\mb,\Vcb)$ and $(\mb, \mupi)$ planes. 
               We compare the results of the two fits including the full sets of measured
               moments, one including hadronic-mass moments (black line) and one including
               moments of the $\nx$ distribution instead (blue dashed line), with
               a fit including only hadronic-mass and lepton-energy moments (red dotted line)
               and a fit including only hadronic-mass moments and partial branching fraction
               measurements (magenta dashed-dotted line). 
               We do not include the additional uncertainty of $1.4\%$
               due to the expansion of $\Gammasl$ in the plotted values of $\Vcb$.
          }
   \label{fig:vcb_chi2_contours}
\end{figure*}

Figure \ref{fig:vcb_chi2_contours} shows $\Delta\chisq = 1$ contours of both 
fits in the $(\mb,\Vcb)$ and $(\mb,\mupi)$ planes. We find an almost identical 
precision for the fit values of $\Vcb$, $\mb$, and $\mupi$. 
In the Figure, we also show the results of two fits with 
reduced sets of input measurements. 
To illustrate the influence of the photon-energy measurements, 
a fit with only hadronic-mass and lepton-energy 
moments is performed. For further comparison we also perform a fit 
with only hadronic-mass moments and partial branching fractions. 
The fits with reduced 
experimental input show a significantly reduced accuracy of the extracted 
parameters.

As our primary results we choose the values extracted from the fit with 
hadronic-mass moments since this fit has been used extensively before.
Its results are in good agreement with earlier determinations 
\cite{Buchmuller:2005globalhqefit, Bauer:2004GlobalFit1SScheme},
but their uncertainties are slightly larger because of the 
restrictions to $\babar$ data only.

The use of combined mass-and-energy moments $\nx$ does not lead to a more 
precise determination of the fundamental physics parameters $\Vcb$, $\mb$, and 
$\mc$. However, the agreement of both fits confirms 
that higher-order corrections, which are needed for the expansion of the 
hadronic-mass moments but not for the $\nx$ moments, have been estimated 
correctly. A significant change in the uncertainties of the SM and HQE parameters 
would have indicated a too naive treatment of the corrections for the mass 
moments \cite{Uraltsev:private}. Consequently, the presented results have increased the 
confidence into the validity of error estimates that have to be made 
for a reliable determination of $\mb$, $\mc$, and $\Vcb$.

\section{Acknowledgments}
We are grateful for the 
extraordinary contributions of our \pep2\ colleagues in
achieving the excellent luminosity and machine conditions
that have made this work possible.
The success of this project also relies critically on the 
expertise and dedication of the computing organizations that 
support \babar.
The collaborating institutions wish to thank 
SLAC for its support and the kind hospitality extended to them. 
This work is supported by the
US Department of Energy
and National Science Foundation, the
Natural Sciences and Engineering Research Council (Canada),
the Commissariat \`a l'Energie Atomique and
Institut National de Physique Nucl\'eaire et de Physique des Particules
(France), the
Bundesministerium f\"ur Bildung und Forschung and
Deutsche Forschungsgemeinschaft
(Germany), the
Istituto Nazionale di Fisica Nucleare (Italy),
the Foundation for Fundamental Research on Matter (The Netherlands),
the Research Council of Norway, the
Ministry of Education and Science of the Russian Federation, 
Ministerio de Educaci\'on y Ciencia (Spain), and the
Science and Technology Facilities Council (United Kingdom).
Individuals have received support from 
the Marie-Curie IEF program (European Union) and
the A. P. Sloan Foundation.

\clearpage

\renewcommand{\thetable}{A.\Roman{table}}
\setcounter{table}{0}


\begin{table*}
\addtolength{\tabcolsep}{1mm}
\caption{Results for the moments $\mxmom{k}$ with $k = 1 \ldots 3$ for different 
         minimum lepton momenta $\plmin$ with statistical and systematic uncertainties.
         The systematic uncertainties are grouped in five categories 
         having related sources: 
         \textsl{MC statistics} contains the statistical uncertainties of the calibration 
         curves and of the residual background.
         \textsl{Simulation related} is the sum of uncertainties due to neutral and charged 
         reconstruction efficiency differences in data and MC, particle identification, and
         mismodeling of final state radiation.
         The category \textsl{extraction method} contains the conservative estimate 
         of half of the bias correction.
         The category \textsl{background} sums all contributions from the variation of 
         the residual background components.
         The category \textsl{signal model} sums the impact of the variation of the 
         signal decay branching fractions.
         Minimum lepton momenta are given in $\gevc$.
         Moments and uncertainties are given in $(\gevcc)^{k}$. \\
        }

\begin{tabular}{lccccccccc}
\hline \hline
$k$ &$ \plmin$ &$\langle m_{X}^{k} \rangle$ &$\sigma_{stat}$ &$\sigma_{sys}$ &MC &simulation &extraction &back- &signal  \\
 & [\gevc]  & & & &statistics &related &method &groud &model  \\
\hline 
1 &0.8 &2.0906 &$\pm 0.0063$ &$\pm 0.0166$ &0.0058 &0.0099 &0.0096 &0.0047 &0.0031  \\
 &0.9 &2.0890 &$\pm 0.0062$ &$\pm 0.0158$ &0.0048 &0.0088 &0.0103 &0.0045 &0.0028  \\
 &1.0 &2.0843 &$\pm 0.0061$ &$\pm 0.0153$ &0.0044 &0.0076 &0.0109 &0.0044 &0.0027  \\
 &1.1 &2.0765 &$\pm 0.0063$ &$\pm 0.0165$ &0.0044 &0.0072 &0.0127 &0.0047 &0.0026  \\
 &1.2 &2.0671 &$\pm 0.0064$ &$\pm 0.0160$ &0.0046 &0.0073 &0.0120 &0.0045 &0.0025  \\
 &1.3 &2.0622 &$\pm 0.0068$ &$\pm 0.0168$ &0.0048 &0.0073 &0.0131 &0.0050 &0.0023  \\
 &1.4 &2.0566 &$\pm 0.0073$ &$\pm 0.0183$ &0.0047 &0.0069 &0.0150 &0.0054 &0.0021  \\
 &1.5 &2.0494 &$\pm 0.0081$ &$\pm 0.0198$ &0.0036 &0.0074 &0.0168 &0.0061 &0.0019  \\
 &1.6 &2.0430 &$\pm 0.0092$ &$\pm 0.0221$ &0.0038 &0.0082 &0.0187 &0.0070 &0.0018  \\
 &1.7 &2.0387 &$\pm 0.0109$ &$\pm 0.0265$ &0.0047 &0.0081 &0.0232 &0.0083 &0.0015  \\
 &1.8 &2.0370 &$\pm 0.0143$ &$\pm 0.0337$ &0.0069 &0.0097 &0.0299 &0.0098 &0.0013  \\
 &1.9 &2.0388 &$\pm 0.0198$ &$\pm 0.0413$ &0.0082 &0.0123 &0.0355 &0.0150 &0.0008  \\
\hline 
2 &0.8 &4.429 &$\pm 0.029$ &$\pm 0.070$ &0.027 &0.047 &0.030 &0.018 &0.008  \\
 &0.9 &4.416 &$\pm 0.027$ &$\pm 0.063$ &0.020 &0.041 &0.033 &0.016 &0.008  \\
 &1.0 &4.394 &$\pm 0.026$ &$\pm 0.058$ &0.020 &0.033 &0.035 &0.015 &0.008  \\
 &1.1 &4.354 &$\pm 0.026$ &$\pm 0.063$ &0.019 &0.031 &0.043 &0.016 &0.008  \\
 &1.2 &4.308 &$\pm 0.026$ &$\pm 0.058$ &0.019 &0.030 &0.039 &0.015 &0.007  \\
 &1.3 &4.281 &$\pm 0.027$ &$\pm 0.061$ &0.020 &0.029 &0.044 &0.016 &0.007  \\
 &1.4 &4.253 &$\pm 0.028$ &$\pm 0.066$ &0.021 &0.028 &0.051 &0.018 &0.006  \\
 &1.5 &4.220 &$\pm 0.031$ &$\pm 0.070$ &0.015 &0.029 &0.058 &0.019 &0.006  \\
 &1.6 &4.183 &$\pm 0.034$ &$\pm 0.078$ &0.015 &0.032 &0.065 &0.022 &0.005  \\
 &1.7 &4.158 &$\pm 0.040$ &$\pm 0.094$ &0.019 &0.032 &0.082 &0.026 &0.004  \\
 &1.8 &4.145 &$\pm 0.051$ &$\pm 0.120$ &0.026 &0.036 &0.107 &0.031 &0.004  \\
 &1.9 &4.136 &$\pm 0.069$ &$\pm 0.142$ &0.031 &0.046 &0.122 &0.048 &0.002  \\
\hline 
3 &0.8 &9.57 &$\pm 0.11$ &$\pm 0.25$ &0.11 &0.18 &0.07 &0.06 &0.02  \\
 &0.9 &9.49 &$\pm 0.10$ &$\pm 0.22$ &0.08 &0.15 &0.08 &0.05 &0.03  \\
 &1.0 &9.41 &$\pm 0.09$ &$\pm 0.18$ &0.06 &0.11 &0.09 &0.04 &0.03  \\
 &1.1 &9.25 &$\pm 0.09$ &$\pm 0.19$ &0.06 &0.10 &0.11 &0.04 &0.03  \\
 &1.2 &9.09 &$\pm 0.08$ &$\pm 0.18$ &0.06 &0.10 &0.10 &0.04 &0.02  \\
 &1.3 &8.98 &$\pm 0.08$ &$\pm 0.18$ &0.06 &0.09 &0.11 &0.04 &0.02  \\
 &1.4 &8.88 &$\pm 0.09$ &$\pm 0.19$ &0.06 &0.09 &0.13 &0.04 &0.02  \\
 &1.5 &8.75 &$\pm 0.09$ &$\pm 0.19$ &0.04 &0.09 &0.15 &0.05 &0.02  \\
 &1.6 &8.61 &$\pm 0.10$ &$\pm 0.22$ &0.05 &0.10 &0.17 &0.06 &0.02  \\
 &1.7 &8.51 &$\pm 0.11$ &$\pm 0.26$ &0.05 &0.10 &0.22 &0.07 &0.01  \\
 &1.8 &8.45 &$\pm 0.14$ &$\pm 0.33$ &0.07 &0.11 &0.29 &0.08 &0.01  \\
 &1.9 &8.38 &$\pm 0.19$ &$\pm 0.38$ &0.08 &0.13 &0.32 &0.12 &0.00  \\
\hline 
\end{tabular} 

\label{tab:massMomentsSummary_1}
\end{table*}

\begin{table*}
\addtolength{\tabcolsep}{1mm}
\caption{Results for the moments $\mxmom{k}$ with $k = 4 \ldots 6$ for different 
         minimum lepton momenta $\plmin$ with statistical and systematic uncertainties.
         The systematic uncertainties are grouped in five categories 
         having related sources: 
         \textsl{MC statistics} contains the statistical uncertainties of the calibration 
         curves and of the residual background.
         \textsl{Simulation related} is the sum of uncertainties due to neutral and charged 
         reconstruction efficiency differences in data and MC, particle identification, and
         mismodeling of final state radiation.
         The category \textsl{extraction method} contains the conservative estimate 
         of half of the bias correction.
         The category \textsl{background} sums all contributions from the variation of 
         the residual background components.
         The category \textsl{signal model} sums the impact of the variation of the 
         signal decay branching fractions.
         moment measurements. Minimum lepton momenta are given in $\gevc$.
         Moments and uncertainties are given in $(\gevcc)^{k}$. \\
        }

\begin{tabular}{lccccccccc}
\hline \hline
$k$ &$\plmin$ &$\langle m_{X}^{k} \rangle$ &$\sigma_{stat}$ &$\sigma_{sys}$ &MC &simulation &extraction &back- &signal  \\
 & [\gevc]  & & & &statistics &related &method &groud &model  \\
\hline 
4 &0.8 &21.20 &$\pm 0.39$ &$\pm 0.84$ &0.35 &0.61 &0.14 &0.19 &0.11  \\
 &0.9 &20.83 &$\pm 0.33$ &$\pm 0.69$ &0.26 &0.51 &0.17 &0.15 &0.11  \\
 &1.0 &20.55 &$\pm 0.30$ &$\pm 0.56$ &0.24 &0.35 &0.19 &0.12 &0.12  \\
 &1.1 &20.01 &$\pm 0.27$ &$\pm 0.55$ &0.19 &0.32 &0.27 &0.11 &0.12  \\
 &1.2 &19.48 &$\pm 0.25$ &$\pm 0.49$ &0.17 &0.29 &0.23 &0.09 &0.10  \\
 &1.3 &19.09 &$\pm 0.25$ &$\pm 0.52$ &0.17 &0.33 &0.27 &0.10 &0.07  \\
 &1.4 &18.77 &$\pm 0.25$ &$\pm 0.52$ &0.17 &0.29 &0.32 &0.11 &0.07  \\
 &1.5 &18.33 &$\pm 0.26$ &$\pm 0.50$ &0.13 &0.24 &0.37 &0.11 &0.06  \\
 &1.6 &17.85 &$\pm 0.27$ &$\pm 0.55$ &0.12 &0.27 &0.42 &0.13 &0.05  \\
 &1.7 &17.50 &$\pm 0.30$ &$\pm 0.66$ &0.14 &0.26 &0.56 &0.15 &0.03  \\
 &1.8 &17.28 &$\pm 0.37$ &$\pm 0.83$ &0.18 &0.27 &0.73 &0.18 &0.03  \\
 &1.9 &16.99 &$\pm 0.48$ &$\pm 0.90$ &0.21 &0.34 &0.76 &0.27 &0.01  \\
\hline 
5 &0.8 &48.51 &$\pm 1.39$ &$\pm 2.90$ &1.37 &2.10 &0.15 &0.64 &0.51  \\
 &0.9 &46.87 &$\pm 1.14$ &$\pm 2.21$ &0.84 &1.67 &0.31 &0.46 &0.49  \\
 &1.0 &46.00 &$\pm 0.97$ &$\pm 1.74$ &0.79 &1.07 &0.36 &0.32 &0.50  \\
 &1.1 &44.20 &$\pm 0.85$ &$\pm 1.61$ &0.57 &0.94 &0.60 &0.30 &0.48  \\
 &1.2 &42.55 &$\pm 0.77$ &$\pm 1.44$ &0.53 &0.88 &0.50 &0.24 &0.37  \\
 &1.3 &41.29 &$\pm 0.72$ &$\pm 1.47$ &0.43 &1.01 &0.61 &0.24 &0.28  \\
 &1.4 &40.31 &$\pm 0.70$ &$\pm 1.45$ &0.47 &0.94 &0.74 &0.25 &0.26  \\
 &1.5 &38.88 &$\pm 0.70$ &$\pm 1.26$ &0.33 &0.65 &0.84 &0.26 &0.23  \\
 &1.6 &37.35 &$\pm 0.70$ &$\pm 1.38$ &0.34 &0.71 &1.00 &0.29 &0.15  \\
 &1.7 &36.28 &$\pm 0.78$ &$\pm 1.61$ &0.34 &0.68 &1.32 &0.34 &0.10  \\
 &1.8 &35.56 &$\pm 0.94$ &$\pm 2.00$ &0.47 &0.69 &1.73 &0.41 &0.08  \\
 &1.9 &34.58 &$\pm 1.18$ &$\pm 2.11$ &0.56 &0.86 &1.73 &0.61 &0.04  \\
\hline 
6 &0.8 &115.20 &$\pm 4.73$ &$\pm 11.43$ &4.39 &6.84 &5.64 &2.02 &3.76  \\
 &0.9 &107.97 &$\pm 3.74$ &$\pm 8.32$ &2.54 &5.36 &3.74 &1.36 &3.22  \\
 &1.0 &105.19 &$\pm 3.09$ &$\pm 6.19$ &2.34 &3.27 &2.26 &0.87 &3.05  \\
 &1.1 &99.35 &$\pm 2.60$ &$\pm 5.19$ &1.90 &2.85 &0.81 &0.79 &2.80  \\
 &1.2 &94.82 &$\pm 2.28$ &$\pm 4.35$ &1.53 &2.49 &0.23 &0.57 &2.16  \\
 &1.3 &91.01 &$\pm 2.05$ &$\pm 4.09$ &1.36 &2.73 &0.31 &0.53 &1.73  \\
 &1.4 &88.02 &$\pm 1.94$ &$\pm 3.86$ &1.23 &2.57 &0.94 &0.53 &1.57  \\
 &1.5 &83.46 &$\pm 1.86$ &$\pm 3.35$ &0.88 &2.20 &1.07 &0.53 &1.47  \\
 &1.6 &78.84 &$\pm 1.81$ &$\pm 3.17$ &0.84 &1.85 &1.73 &0.61 &0.97  \\
 &1.7 &75.87 &$\pm 1.98$ &$\pm 3.92$ &0.91 &1.73 &3.10 &0.76 &0.29  \\
 &1.8 &73.66 &$\pm 2.35$ &$\pm 4.70$ &1.06 &1.69 &4.05 &0.91 &0.25  \\
 &1.9 &70.70 &$\pm 2.83$ &$\pm 4.77$ &1.23 &2.09 &3.86 &1.33 &0.14  \\
\hline 
\hline 
\end{tabular} 

\label{tab:massMomentsSummary_2}
\end{table*}


\begin{table*}
\caption{Results for \moment{\nxn}  for $k=2,4,6$ 
for all minimum lepton momentum values $\plmin$.
The statistical uncertainty contains the uncertainty arising from the limited data sample 
and an additional statistical uncertainty arising from the determination of the combinatorial 
background. The systematic uncertainties are grouped in five categories 
having related sources: 
\textsl{MC statistics} contains the statistical uncertainties of the calibration curves and of the
residual background.
\textsl{Simulation related} is the sum of 
neutral and charged reconstruction efficiency differences in data and MC, 
\epmiss differences, mismodeling of final state radiation, and PID impact.
The category \textsl{extraction method} contains 
the conservative estimate of half of the bias correction
and the impact of the calibration curve binning. 
The category \textsl{background} sums all contributions from 
the variation of the residual background components.
The category \textsl{signal model} sums the impact of the variation of the signal decay branching
fractions. \\
 }

\begin{tabular}{lcrcrcrccccc}
\toprule
$k$ &\plmin &\moment{\nxn} & &$\sigma_\mathrm{stat}$ & &$\sigma_\mathrm{sys}$ &MC &simulation &extraction &back- &signal  \\
 &  [\gevc] & & & & & &statistics &related &method &groud &model  \\
\hline 
2 &0.8 &1.522 &$\pm$ &0.049 &$\pm$ &0.056 &0.020 &0.050 &0.011 &0.012 &0.004  \\
 &0.9 &1.483 &$\pm$ &0.047 &$\pm$ &0.057 &0.015 &0.054 &0.009 &0.009 &0.004  \\
 &1.0 &1.465 &$\pm$ &0.044 &$\pm$ &0.041 &0.013 &0.037 &0.009 &0.008 &0.003  \\
 &1.1 &1.438 &$\pm$ &0.037 &$\pm$ &0.040 &0.012 &0.037 &0.009 &0.006 &0.003  \\
 &1.2 &1.449 &$\pm$ &0.034 &$\pm$ &0.038 &0.011 &0.036 &0.006 &0.005 &0.002  \\
 &1.3 &1.428 &$\pm$ &0.031 &$\pm$ &0.031 &0.010 &0.027 &0.006 &0.006 &0.004  \\
 &1.4 &1.400 &$\pm$ &0.030 &$\pm$ &0.028 &0.009 &0.025 &0.006 &0.006 &0.004  \\
 &1.5 &1.369 &$\pm$ &0.035 &$\pm$ &0.032 &0.009 &0.029 &0.008 &0.007 &0.005  \\
 &1.6 &1.346 &$\pm$ &0.033 &$\pm$ &0.027 &0.009 &0.020 &0.014 &0.007 &0.003  \\
 &1.7 &1.344 &$\pm$ &0.037 &$\pm$ &0.029 &0.010 &0.020 &0.015 &0.008 &0.004  \\
 &1.8 &1.337 &$\pm$ &0.038 &$\pm$ &0.035 &0.013 &0.014 &0.029 &0.008 &0.004  \\
 &1.9 &1.196 &$\pm$ &0.032 &$\pm$ &0.033 &0.017 &0.017 &0.020 &0.009 &0.003  \\
\hline 
4 &0.8 &3.54 &$\pm$ &0.41 &$\pm$ &0.39 &0.14 &0.34 &0.08 &0.10 &0.03  \\
 &0.9 &3.21 &$\pm$ &0.37 &$\pm$ &0.36 &0.11 &0.32 &0.09 &0.05 &0.02  \\
 &1.0 &3.00 &$\pm$ &0.29 &$\pm$ &0.25 &0.09 &0.21 &0.09 &0.04 &0.02  \\
 &1.1 &2.74 &$\pm$ &0.22 &$\pm$ &0.17 &0.06 &0.14 &0.09 &0.01 &0.02  \\
 &1.2 &2.81 &$\pm$ &0.19 &$\pm$ &0.20 &0.06 &0.15 &0.12 &0.02 &0.03  \\
 &1.3 &2.60 &$\pm$ &0.15 &$\pm$ &0.16 &0.04 &0.10 &0.11 &0.01 &0.04  \\
 &1.4 &2.51 &$\pm$ &0.13 &$\pm$ &0.12 &0.04 &0.08 &0.09 &0.01 &0.03  \\
 &1.5 &2.34 &$\pm$ &0.13 &$\pm$ &0.13 &0.03 &0.09 &0.09 &0.00 &0.02  \\
 &1.6 &2.11 &$\pm$ &0.10 &$\pm$ &0.09 &0.03 &0.06 &0.06 &0.00 &0.01  \\
 &1.7 &2.03 &$\pm$ &0.12 &$\pm$ &0.08 &0.03 &0.06 &0.04 &0.00 &0.01  \\
 &1.8 &1.98 &$\pm$ &0.10 &$\pm$ &0.06 &0.04 &0.04 &0.02 &0.00 &0.01  \\
 &1.9 &1.57 &$\pm$ &0.07 &$\pm$ &0.06 &0.03 &0.05 &0.02 &0.00 &0.01  \\
\hline 
6 &0.8 &13.52 &$\pm$ &3.93 &$\pm$ &3.42 &1.37 &2.97 &0.49 &0.81 &0.34  \\
 &0.9 &10.87 &$\pm$ &2.78 &$\pm$ &2.65 &0.93 &2.39 &0.52 &0.37 &0.24  \\
 &1.0 &9.02 &$\pm$ &2.22 &$\pm$ &1.88 &0.84 &1.55 &0.54 &0.29 &0.20  \\
 &1.1 &7.06 &$\pm$ &1.35 &$\pm$ &0.78 &0.35 &0.58 &0.34 &0.07 &0.14  \\
 &1.2 &7.50 &$\pm$ &1.16 &$\pm$ &0.92 &0.32 &0.68 &0.49 &0.11 &0.18  \\
 &1.3 &6.28 &$\pm$ &0.84 &$\pm$ &0.64 &0.22 &0.38 &0.41 &0.06 &0.20  \\
 &1.4 &5.83 &$\pm$ &0.62 &$\pm$ &0.49 &0.16 &0.30 &0.32 &0.06 &0.12  \\
 &1.5 &4.99 &$\pm$ &0.49 &$\pm$ &0.52 &0.13 &0.30 &0.40 &0.03 &0.05  \\
 &1.6 &3.93 &$\pm$ &0.32 &$\pm$ &0.31 &0.11 &0.16 &0.24 &0.03 &0.03  \\
 &1.7 &3.63 &$\pm$ &0.35 &$\pm$ &0.22 &0.09 &0.13 &0.15 &0.02 &0.03  \\
 &1.8 &3.42 &$\pm$ &0.23 &$\pm$ &0.19 &0.09 &0.11 &0.12 &0.02 &0.03  \\
 &1.9 &2.51 &$\pm$ &0.16 &$\pm$ &0.13 &0.06 &0.11 &0.03 &0.02 &0.02  \\
\hline \hline
\end{tabular} 

\label{tab:NxOrder246}
\end{table*}


\begin{thebibliography}{59}
\bibitem[{Cha()}]{ChargeConjugation}
\bibinfo{title}{Charge-conjugate processes are implied throughout this paper.}
\end{thebibliography}
\end{document}